\documentclass[12pt]{iopart}

\usepackage{iopams,graphicx}

\begin{document}

\title[Lasing and antibunching of optical phonons in semiconductor DQDs]
{Lasing and antibunching of optical phonons in
semiconductor double quantum dots}

\author{R Okuyama$^1$, M Eto$^1$, and T Brandes$^2$}
\address{$^1$ Faculty of Science and Technology,
    Keio University, Yokohama 223-8522, Japan}
\address{$^2$ Institut f\"ur Theoretische Physik,
    Technische Universit\"at Berlin, D-10623 Berlin, Germany}
\ead{rokuyama@rk.phys.keio.ac.jp}

\begin{abstract}
We theoretically propose optical phonon lasing in a double quantum dot
(DQD) fabricated on a semiconductor substrate.
No additional cavity or resonator is required.
An electron in the DQD is found to be coupled to only two
longitudinal optical phonon modes
that act as a natural cavity.
When the energy level spacing in the DQD is tuned to the phonon energy,
the electron transfer is accompanied by the emission of the phonon modes.
The resulting non-equilibrium motion of electrons and phonons
is analyzed by the rate equation approach based on
the Born-Markov-Secular approximation.
We show that the lasing occurs for pumping the DQD via electron tunneling
at rate much larger than the phonon decay rate,
whereas a phonon antibunching is observed in the opposite regime of
slow tunneling.
Both effects disappear by an effective thermalization induced by
the Franck-Condon effect in a DQD fabricated in a suspended
carbon nanotube with strong electron-phonon coupling.
\end{abstract}


\maketitle

\section{Introduction}

In conventional lasers,
two-level systems couple to a single mode of photon in a cavity.
The pumping of electrons to the upper level results in a
light amplification through the stimulated emission of radiation.
Recently, the lasing was reported for a single atom in a cavity,
which is called microlaser \cite{McKeever2003}.
Such a system is being intensively studied in the context of cavity
quantum electrodynamics (QED)
    \cite{Walther2006},
which also works as a single photon source
to produce antibunched photons
    \cite{Choi2006}.

Quantum dots are electrically tunable two-level systems.
The cavity QED using a quantum dot has a potential for wider
application to the quantum information processing
    \cite{Hennessy2007}
as well as the single photon source
    \cite{Bozyigit2011}.
When the quantum dot is connected
to an external circuit, the electronic state in the quantum dot
can be controlled by the electric current.
The microlaser was realized in the so-called circuit QED,
in which a superconducting quantum dot in a circuit is
coupled to a microwave resonator \cite{Astafiev2007}.
In this case, the pumping is realized using the superconducting
circuit. The electric current drives the lasing when the
level spacing is tuned to the microwave energy
   \cite{Blais2004,Ashhab2009,Andre2009,Gartner2011,Xiang2013}.

In the present work, we theoretically examine the transport
through a semiconductor double quantum dot (DQD) in the presence
of electron-optical-phonon coupling
and propose a phonon lasing without a cavity or resonator.
The electron-phonon interaction in quantum dots reveals itself in the
transport phenomena, which was investigated in various contexts
until now.
For DQDs fabricated in InAs nanowire and graphene, an interference
pattern of electric current was observed as a function of level spacing
in the DQDs, which is ascribable to the emission of acoustic phonons
    \cite{Roulleau2011}.
It is the Dicke-type interference between two transport processes in
which a longitudinal acoustic (LA) phonon is emitted in one dot or another
    \cite{Brandes1999}.
In a single quantum dot fabricated in a suspended carbon nanotube
(CNT), the Franck-Condon blockade was reported
    \cite{Sapmaz2006,Leturcq2009}.
Due to the strong electron-phonon interaction in the CNT,
the electric transport is accompanied by the lattice distortion,
which results in the current
suppression under a small bias voltage
    \cite{Koch2005}.
This is the manifestation of the Franck-Condon effect
in the electric transport, which was originally known in the optical
absorption of molecules
    \cite{Atkins}.
Regarding the study of optical phonons,
Amaha and Ono observed the phonon-assisted transport through a DQD.
The current is markedly enhanced when the level spacing in the DQD is
tuned to an integer multiple of the energy of longitudinal optical (LO)
phonons in the semiconductor substrate
    \cite{Amaha2012}.

In this paper, we show the LO-phonon lasing in the phonon-assisted
transport through a DQD.
First, we show that a DQD effectively couples to only two LO phonon modes.
The phonon modes do not diffuse and act as a natural cavity since the
optical phonons have a flat dispersion relation. Thus
our laser does not require a cavity or resonator.
The pumping to the upper level is realized by an electric current
through the DQD under a finite bias voltage, in a similar manner to
the microlaser in the circuit QED \cite{Astafiev2007}.
Thus the pumping rate is determined by the tunneling rate between the
DQD and leads, $\Gamma_{L,R}$.
As we discuss in section \ref{sec:discussion},
the amplified LO phonons occasionally escape from the ``cavity''
by decaying into so-called daughter phonons
    \cite{Vallee1994}
that can be observed externally.
When the pumping rate $\Gamma_{L,R}$ is much larger than the phonon
decay rate $\Gamma_{\rm ph}$, the stimulated emission of phonons,
i.e., phonon lasing, takes place.
We proposed a basic idea of the optical phonon lasing in our previous
letter
    \cite{Okuyama2013}.
In this paper, we present further comprehensive discussion on
the phonon lasing and address the possible experimental realization.

We also find the phonon antibunching in the same system if the
pumping rate $\Gamma_{L,R}$ is smaller than $\Gamma_{\rm ph}$.
In this situation, the phonon emission is regularized by the
single electron transport through the DQD.
We emphasize that the phonon statistics can be changed by
electrically tuning the tunnel coupling between DQD and leads.
Note that LO-phonon-assisted transport through a DQD was
theoretically studied by Gnodtke \etal
    \cite{Gnodtke2006}.
We also note that lasing for acoustic phonons was
demonstrated using phonon-assisted transport through
quantum wells fabricated in a semiconductor superlattice,
which works as a cavity for acoustic modes
    \cite{Beardsley2010}.
Acoustic phonon lasers by optical pumping
    \cite{Kabuss2012}
and transport in the spin blockade regime
    \cite{Khaetskii2013}
were proposed in a single quantum dot embedded in a superlattice.

The electron-phonon coupling in DQDs fabricated in CNTs
is much stronger than the electron-optical-phonon coupling in
DQDs made in GaAs substrate, as we discuss in section \ref{sec:model}.
Both the phonon lasing and antibunching are 
spoilt by phonon thermalization via the Franck-Condon effect
in the former case.
In the electric transport, the number of electrons
in the DQD fluctuates, which is accompanied by the lattice distortion
and thus the creation of bunched phonons.
We show that this effect is negligible in weak coupling case of
semiconductor-based DQDs but surpasses
the lasing and antibunching in strong coupling case of
CNTs\footnote{
    The coupling to photons in a cavity corresponds to the weak
    coupling case, with dimensionless coupling constant
    $\lambda \sim 10^{-4}$ in
        \cite{McKeever2003}
    and $10^{-2}$ in
        \cite{Astafiev2007},
    in equation (\ref{eq:H_eff}).
}.
We also show that the strong electron-phonon coupling brings about
the Franck-Condon blockade in a DQD with finite bias voltages,
as in the case of single quantum dots
    \cite{Sapmaz2006,Leturcq2009,Koch2005}.

This paper is organized as follows.
In section \ref{sec:model}, we explain our model and calculation
method. Starting from the microscopic electron-optical-phonon interaction,
we show that only two phonon modes, $S$- and $A$-phonons, are
coupled to an electron in the DQD.
The effective Hamiltonian is then derived in terms of the phonon modes.
Based on the Born-Markov-Secular approximation,
we obtain the rate equation for the non-equilibrium dynamics
of electrons and phonons in the DQD under a finite bias.
In section \ref{sec:A-phonon},
we take into account $A$-phonons and disregard $S$-phonons.
We examine the electron transport accompanied by the phonon emission.
This results in the phonon lasing or antibunching in the weak
coupling case, whereas it brings about the phonon thermalization
in the strong coupling case.
These different situations are elucidated by the analytical solution of the
rate equation as well as the numerical studies.
In section \ref{sec:S-phonon},
$S$-phonons are examined without $A$-phonons.
$S$-phonons do not contribute to the phonon-assisted
tunneling in the DQD, in contrast to $A$-phonons, and hence
they are irrelevant to the phonon lasing and antibunching.
We examine the Franck-Condon blockade under a finite bias
by the coupling to $S$-phonons as well as $A$-phonons.
Section \ref{sec:AS-phonon} is devoted to the investigation of
general situations in the presence of both $A$- and
$S$-phonons. We show that $S$-phonons do not disturb the lasing or
antibunching of $A$-phonons.
In section \ref{sec:discussion},
we discuss the validity of our theory and 
address possible experimental realizations to observe the phonon
lasing and antibunching.
Finally, we present our conclusions in section \ref{sec:conclusion}.

\begin{figure}
\begin{center}
    \includegraphics[width=12cm]{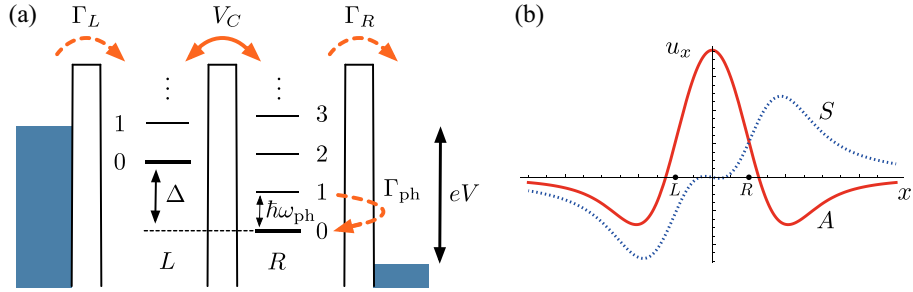}
    \caption{
    (a) Model for a double quantum dot (DQD) coupled to LO
    phonons. The bias voltage $V$ is applied between external leads.
    The spacing $\Delta$ between
    the energy levels in dots $L$ and $R$ is electrically tunable.
    When $\Delta$ matches an integer ($\nu$) multiple of the phonon
    energy $\hbar \omega_{\rm ph}$, the electronic state $|L \rangle_e$ with
    $n$ phonons is coherently coupled to $|R \rangle_e$ with
    $(n+\nu)$ phonons.
    (b) The phonon mode functions ${\bi u}_{S,A}({\bi r})$
    along a line through the centers of quantum dots
    located at $x=\pm \mathcal{R}$,
    when the electron distributions,
    $|\psi_L (\bi{r})|^2$ and $|\psi_R (\bi{r})|^2$, are spherical
    with radius $\mathcal{R}$.
    The $x$-component of ${\bi u}(x,0,0)$ is shown for
    $S$($A$)-phonons which couple (anti-)symmetrically to the DQD.
    Note that $u_x$ is an odd (even) function of $x$
    for $S$($A$)-phonons since the induced charge is proportional to
    $\nabla \cdot \bi{u}(\bi{r})$.
    \label{fig:model}}
\end{center}
\end{figure}

\section{Model and calculation method
    \label{sec:model}}

\subsection{Phonon modes coupled to DQD and effective Hamiltonian}

Figure \ref{fig:model}(a) depicts our model of a DQD embedded
in a semiconductor substrate,
in which two single-level quantum dots, $L$ and $R$, are
connected by tunnel coupling $V_{\rm C}$.
The energy levels, $\varepsilon_L$ and $\varepsilon_R$, are
electrically tunable. We choose $\varepsilon_R=-\varepsilon_L$
and denote the level spacing $\varepsilon_L-\varepsilon_R$
by $\Delta$. We assume that the total number of electrons
in the DQD is restricted to one or zero
due to the Coulomb blockade. The electron couples to
LO phonons of energy $\hbar \omega_{\rm ph}$ in the substrate
by the Fr\"ohlich interaction. Our system Hamiltonian is
$
    \mathcal{H} = \mathcal{H}_e + \mathcal{H}_{\rm ph}
        + \mathcal{H}_{\rm ep}
$,
\begin{eqnarray}
    \mathcal{H}_e &= \frac{\Delta}{2} (n_L - n_R)
        + V_{\rm C} (d_L^\dagger d_R + d_R^\dagger d_L), \\
    \mathcal{H}_{\rm ph} &= \hbar \omega_{\rm ph} \sum_{\bi{q}} N_{\bi{q}},
        \\
    \mathcal{H}_{\rm ep} &=  \sum_{\alpha = L, R} \sum_{\bi{q}}
       M_{\alpha, \bi{q}} (a_{\bi{q}} + a_{-\bi{q}}^\dagger) n_\alpha,
        \label{eq:H_ep}
\end{eqnarray}
using creation (annihilation) operators
$d_{\alpha}^{\dagger}$ ($d_{\alpha}$) for an electron in
dot $\alpha$ and $a_{\bi{q}}^\dagger$ ($a_{\bi{q}}$) for
a phonon with wavevector $\bi{q}$.
$n_{\alpha} = d_{\alpha}^\dagger d_{\alpha}$ and
$N_{\bi{q}} = a_{\bi{q}}^\dagger a_{\bi{q}}$ are the
number operators.
The spin index is omitted for electrons.
The coupling constant is given by
\begin{eqnarray}
    M_{\alpha, \bi{q}} =
    \sqrt{\frac{\hbar \omega_{\rm ph} e^2}{2 \mathcal{V}} \left[
        \frac{1}{\epsilon(\infty)} - \frac{1}{\epsilon(0)}
    \right]} \frac{1}{q}
        \int \rmd\bi{r} |\psi_{\alpha} (\bi{r})|^2
        \rme^{\rmi \bi{q} \cdot \bi{r}},
    \label{eq:M}
\end{eqnarray}
where $\epsilon(\infty)$ [$\epsilon(0)$] is the dielectric constant
at high [low] frequency, $\mathcal{V}$ is the volume of substrate,
and $\psi_{\alpha} (\bi{r})$ is
the electron wavefunction in dot $\alpha$ of radius $\mathcal{R}$.
The LO phonons only around the $\Gamma$ point, such as
$|\bi{q}| \lesssim 1/\mathcal{R}$, are coupled to the DQD
because of an oscillating factor in the integral over
$|\psi_{\alpha} (\bi{r})|^2$. This fact justifies the dispersionless
phonons in $\mathcal{H}_{\rm ph}$.
We assume equivalent quantum dots $L$ and $R$, whence
$M_{R, \bi{q}}=M_{L, \bi{q}} \rme^{\rmi \bi{q} \cdot \bi{r}_{LR}}$ with
$\bi{r}_{LR}$ being a vector joining their centers.

In $\mathcal{H}_{\rm ep}$,
an electron in dot $\alpha$ couples to a single mode of phonon described by
\begin{eqnarray}
    a_\alpha = \frac{\sum_{\bi{q}} M_{\alpha, \bi{q}} a_{\bi{q}}}
        {(\sum_{\bi{q}} |M_{\alpha, \bi{q}}|^2)^{1/2}}.
   \label{eq:a_La_R}
\end{eqnarray}
We perform a unitary transformation for phonons from $a_{\bi{q}}$
to $S$- and $A$-phonon modes,
\begin{eqnarray}
    a_{S} = \frac{a_{L} + a_{R}}{\sqrt{2 + (\mathcal{S} + \mathcal{S}^*)}},
\quad
    a_{A} = \frac{a_{L} - a_{R}}{\sqrt{2 - (\mathcal{S} + \mathcal{S}^*)}},
    \label{eq:a_Sa_A}
\end{eqnarray}
and others orthogonal to $a_S$ and $a_A$, where
$\mathcal{S}$
is the overlap integral between $a_L$ and $a_R$ phonons
in equation (\ref{eq:a_La_R})\footnote{
    The overlap integral,
    $
        \mathcal{S} = _{\rm ph}\langle L|R \rangle_{\rm ph}
        = _{\rm ph}\langle 0|a_L a_R^\dagger|0 \rangle_{\rm ph}
        = [a_L, a_R^\dagger]
    $,
    is evaluated in \ref{app:mode_function}.
}.
Disregarding the modes decoupled from the DQD,
we obtain the effective Hamiltonian
\begin{eqnarray}
    H=\mathcal{H}_e &+
      \hbar \omega_{\rm ph}
        \left[ N_S + \lambda_S (a_S+ a_S^\dagger)
                            (n_L + n_R) \right]
      \nonumber \\
      &+ \hbar \omega_{\rm ph}
        \left[ N_A + \lambda_A (a_A+ a_A^\dagger)
                            (n_L - n_R) \right],
        \label{eq:H_eff}
\end{eqnarray}
where $N_S = a_{S}^\dagger a_{S}$ and $N_A = a_{A}^\dagger a_{A}$,
with dimensionless coupling constants
\begin{equation}
    \lambda_{S/A} = \frac{1}{2 \hbar \omega_{\rm ph}}
    \left( \sum_{\bi{q}}
    |M_{L, \bi{q}} \pm M_{R, \bi{q}}|^2 \right)^{1/2}.
    \label{eq:lambda}
\end{equation}

The mode functions for $S$- and $A$-phonons are shown in figure
\ref{fig:model}(b)
along a line through the centers of the quantum dots.
The definition and calculation of the mode functions are given in
\ref{app:mode_function}.
Since the phonons are dispersionless, they
do not diffuse and act as a cavity including the DQD\footnote{
    If a weak quadratic dispersion around the $\Gamma$ point is taken
    into account, the phonon modes are scattered by the rate of
    $
        \partial^2 \omega_{\rm ph} /
        \partial q^2 |_{q = 0} / \mathcal{R}^2
    $,
    which is smaller than the decay rate $\Gamma_{\rm ph}$ of
    LO phonons by two orders of magnitude in GaAs quantum dots
    with $\mathcal{R} = 10 \sim 100~{\rm nm}$.
}.
$A$-phonons play a crucial role in the phonon-assisted tunneling
between the quantum dots and thus in the phonon lasing,
as discussed below,
whereas $S$-phonons do not since it couples to the {\em total}
number of electrons in the DQD, $n_L + n_R$.
Both phonons are relevant to the Franck-Condon effect.

Our Hamiltonian $H$ in equation (\ref{eq:H_eff})
is applicable to  DQDs fabricated in a semiconductor substrate,
where $\hbar \omega_{\rm ph} = 36~{\rm meV}$
and $\lambda_{S,A} = 0.1 \sim 0.01$ for
$\mathcal{R} = 10 \sim 100~{\rm nm}$ in GaAs.
It also describes a DQD in a suspended
CNT when an electron couples to a vibron, longitudinal stretching mode
with $\hbar \omega_{\rm ph} \sim 1~{\rm meV}$, $\lambda_A \gtrsim 1$,
and $\lambda_S = 0$ in experimental situations
    \cite{Sapmaz2006, Leturcq2009}, as shown in \ref{app:CNT}.

\subsection{Rate equation in energy eigenbasis}

The DQD is connected to external leads $L$ and $R$ in series,
which enables the electronic pumping by the electric current
under a finite bias.
The tunnel coupling between lead $L$ and dot $L$ is denoted by
$\Gamma_{L}$ and that between lead $R$ and dot $R$ is by $\Gamma_{R}$.
We also introduce the phonon decay rate $\Gamma_{\rm ph}$ to
take into account a natural decay of LO phonons into so-called
daughter phonons due to the lattice anharmonicity
    \cite{Vallee1994}.
We describe the dynamics of the DQD-phonon density matrix $\rho$
using a Markovian master equation
\begin{eqnarray}
    \dot \rho &= - \frac{\rmi}{\hbar} [H, \rho]
        + \mathcal{L}_e \rho + \mathcal{L}_{\rm ph} \rho,
        \label{eq:master}
\end{eqnarray}
where $\mathcal{L}_e$ and $\mathcal{L}_{\rm ph}$
describe the electron tunneling between the DQD and leads and
the phonon decay, respectively.
$\mathcal{L}_e$ is written as
\begin{eqnarray}
\fl \mathcal{L}_e \rho = \sum_{\alpha = L, R; i, j}
    \frac{\Gamma_\alpha}{2}
&\Bigl[ f_\alpha(\epsilon_i - \epsilon_j) \Bigl(
    |i \rangle \langle i| d_\alpha^\dagger |j \rangle
    \langle j| \rho d_\alpha
  + d_\alpha^\dagger \rho |j \rangle \langle j| d_\alpha
 |i \rangle \langle i|
 \nonumber \\ & \qquad \qquad \qquad
  - \rho |j \rangle \langle j| d_\alpha |i \rangle \langle i|d_\alpha^\dagger
  - d_\alpha |i \rangle \langle i| d_\alpha^\dagger |j \rangle \langle j| \rho
    \Bigr)
\nonumber \\
    &+ \bar{f}_\alpha(\epsilon_i - \epsilon_j)
    \Bigl(
    |j \rangle \langle j| d_\alpha |i \rangle \langle i| \rho d_\alpha^\dagger
  + d_\alpha \rho |i \rangle \langle i| d_\alpha^\dagger |j \rangle \langle j|
  \nonumber \\ & \qquad \qquad \qquad
  - \rho |i \rangle \langle i| d_\alpha^\dagger |j \rangle \langle j| d_\alpha
  - d_\alpha^\dagger |j \rangle \langle j| d_\alpha |i \rangle \langle i| \rho
  \Bigr) \Bigr],
  \label{eq:L_e}
\end{eqnarray}
with $|i \rangle$ and $\epsilon_i$ being an eigenstate of $H$
and the corresponding energy eigenvalue, respectively, and
$f_\alpha (\epsilon)$ $[\bar{f}_\alpha (\epsilon) = 1 - f_\alpha (\epsilon)]$
being the Fermi distribution function for electrons [holes] in lead $\alpha$
    \cite{Hubener2009}.
The Fermi levels in leads $L$ and $R$
are given by $\mu_L = eV / 2$ and $\mu_R = - eV / 2$, respectively,
with bias voltage $V$ between the leads.

In the limit of large bias voltage, $\mathcal{L}_e$ is reduced to
\begin{eqnarray}
    \mathcal{L}_e \rho = ( \Gamma_L \mathcal{D} [d_L^\dagger]
    + \Gamma_R \mathcal{D} [d_R] ) \rho,
    \label{eq:L_e_large_V}
\end{eqnarray}
where $\mathcal{D}[A] \rho = A \rho A^\dagger
- \frac{1}{2} \{ \rho, A^\dagger A \}$ is a Lindblad dissipator.
In this case, an electron tunnels into dot $L$ from lead $L$ with
tunneling rate $\Gamma_{L}$ and tunnels out from dot $R$ to
lead $R$ with $\Gamma_{R}$ in one direction.
We examine this situation in the main part of this paper.
With finite bias voltages, equation (\ref{eq:L_e}) is
evaluated in sections \ref{subsec:FCB_A} and \ref{subsec:FCB_S},
where the electron tunneling
takes place in both directions unless $eV$ is far beyond the
temperature $T$.

The phonon dissipator $\mathcal{L}_{\rm ph}$ is given by
\begin{eqnarray}
    \mathcal{L}_{\rm ph} \rho = \Gamma_{\rm ph}
    (\mathcal{D} [a_S] + \mathcal{D} [a_A]) \rho,
\end{eqnarray}
on the assumption that the temperature $T$ in the substrate is
much smaller than $\hbar \omega_{\rm ph}$
and daughter phonons immediately escape from the surroundings of
the DQD.

In the following,
we adopt the Born-Markov-Secular (BMS) approximation
    \cite{Breuer}
to equation (\ref{eq:master}).
We diagonalize the Hamiltonian $H$ in equation (\ref{eq:H_eff})
and set up the rate equation in the energy eigenbasis,
\begin{eqnarray}
    \dot{P_i} &= \sum_{j} L_{ij} P_j
    \label{eq:rate}
\end{eqnarray}
for the probability $P_i$ to find the system
in eigenstate $|i\rangle$. Here,
$
    L_{ij} = \langle i| [
        (\mathcal{L}_e + \mathcal{L}_{\rm ph}) |j \rangle \langle j|
    ] |i \rangle.
$
The solution of equation (\ref{eq:rate}) with $\dot{P_i}=0$
determines the steady state properties. The condition to justify
the BMS approximation will be given in section \ref{sec:discussion}.

\section{Lasing and antibunching of $A$-phonons}
    \label{sec:A-phonon}

In this section, we examine $A$-phonons,
disregarding $S$-phonons by fixing at $\lambda_S = 0$.
The results in this section
are not modified by a finite coupling to
$S$-phonons, as seen in section \ref{sec:AS-phonon}.

\begin{figure}
\begin{center}
    \includegraphics[width=6.5cm]{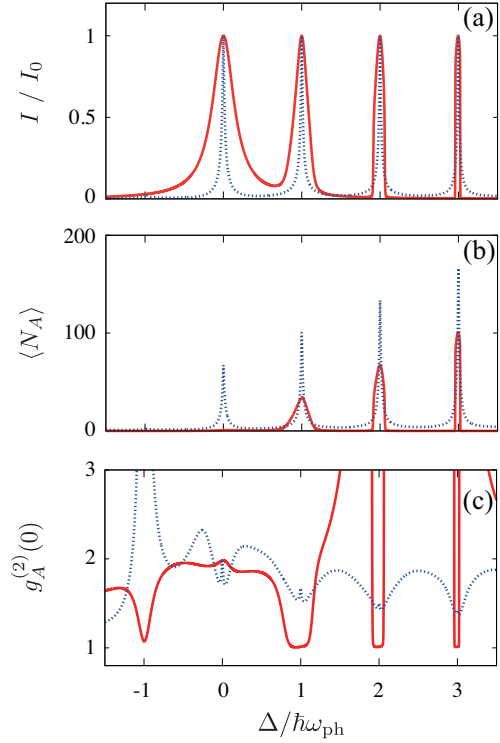}
    \caption{
    (a) Electric current through the DQD,
    (b) $A$-phonon number $\langle N_A \rangle$,
    and (c) its autocorrelation function
    $g^{(2)}_A (0)$ in the large bias-voltage limit,
    as a function of level spacing $\Delta$ in the DQD.
    The dimensionless electron-phonon coupling constants are
    $\lambda_A = 0.1$ (solid lines) or $1$ (dotted lines),
    and $\lambda_S = 0$. In panel (a),
    $I_0 = e \Gamma_R / (2 + \Gamma_R / \Gamma_L)$ is the current
    at $\Delta=0$ in the absence of electron-phonon coupling.
    $\Gamma_L = \Gamma_R = 100~\Gamma_{\rm ph}$
    and $V_{\rm C} = 0.1~\hbar \omega_{\rm ph}$.
    \label{fig:Delta}}
\end{center}
\end{figure}

\subsection{Phonon-assisted transport and phonon lasing
    \label{subsec:lasing}}

First, we present our numerical results in the case of
$\Gamma_{L, R} \gg \Gamma_{\rm ph}$.
We consider the limit of large bias voltage.
Figure \ref{fig:Delta}(a) shows the current $I$ through the DQD as
a function of level spacing $\Delta$, with $\lambda_A=0.1$ (solid line)
and $1$ (dotted line).
Beside the main peak at $\Delta = 0$, we observe
subpeaks at $\Delta = \Delta_{\nu} \simeq \nu \hbar \omega_{\rm ph}$
($\nu = 1, 2, 3, \ldots$) due to the phonon-assisted
tunneling\footnote{
    $\Delta_{\nu}$ is not exactly
    equal to $\nu \hbar \omega_{\rm ph}$ in the presence of tunnel coupling
    $V_{\rm C}$ between the quantum dots.
}.
At the $\nu$th subpeak, electron transport through
the DQD is accompanied by the emission of $\nu$ phonons.
As a result, the phonon number is markedly enhanced at the
subpeaks, as shown in figure \ref{fig:Delta}(b), in both cases of
$\lambda_A=0.1$ and $1$. However, the physics is very different for
the two cases, as we will show below.

For $\lambda_A=0.1$ and $\Delta = \Delta_{\nu}$,
the electronic state
$| L \rangle_e$ with $n$ phonons is coherently coupled to
$| R \rangle_e$ with $(n+\nu)$ phonons
    \cite{Hameau1999},
similarly to cavity QED systems,
if the lattice distortion is neglected.
To examine the amplification of $A$-phonons,
we calculate the phonon autocorrelation function
\begin{eqnarray}
    g^{(2)}_A (\tau) = \frac{\langle :N_A (0) N_A (\tau): \rangle}
    {\langle N_A \rangle^2}.
    \label{eq:g2func}
\end{eqnarray}
The numerator includes the normal product,
$:N_A (0) N_A (\tau):=a_{A}^\dagger(0) a_{A}^\dagger(\tau)
a_{A}(\tau) a_{A}(0)$.
$g^{(2)}_A (\tau)$ is proportional to the probability of
phonon emission at time $\tau$ on the condition that a phonon is
emitted at time $0$
    \cite{Scully, Emary2012}.
A value of $g^{(2)}_A (0)=1$ indicates a
{\em Poisson} distribution of phonons
which is a criterion of phonon lasing, whereas $g^{(2)}_A (0)<1$
[$g^{(2)}_A (0)>1$] represents the phonon antibunching [bunching].
We thus find phonon lasing at the current subpeaks
in figure \ref{fig:Delta}(c) in the case of $\lambda_A=0.1$
(solid line).

When $\lambda_A=1$, the strength of the
electron-phonon interaction is comparable to the phonon energy.
In this case, the
lattice distortion by the Franck-Condon effect seriously disturbs
the above-mentioned coherent coupling between an electron and phonons
in the DQD and, as a result, suppresses the phonon lasing.
Indeed, $g^{(2)}_A (0)>1$ at the current subpeaks, indicating
the phonon bunching.

\begin{figure}
\begin{center}
    \includegraphics[width=8cm]{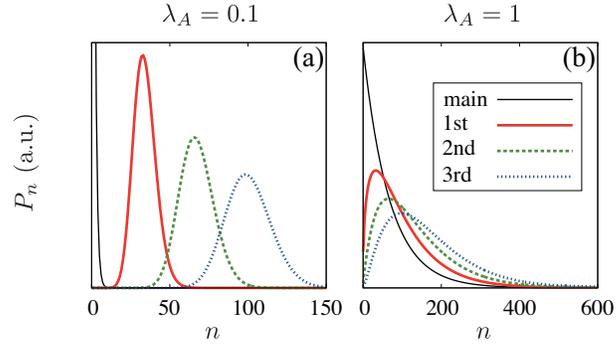}
    \caption{
    Number distribution
    of $A$-phonons at the current main peak ($\Delta = 0$) and
    subpeaks ($\Delta = \Delta_{\nu} \simeq \nu \hbar \omega_{\rm ph}$,
    $\nu=1,2,3$) in figure \ref{fig:Delta}(a).
    (a) $\lambda_A = 0.1$ or (b) 1, and $\lambda_S = 0$.
    The other parameters are the same as in figure \ref{fig:Delta}.
    \label{fig:distribution}}
\end{center}
\end{figure}
\begin{figure}
\begin{center}
    \includegraphics[width=8cm]{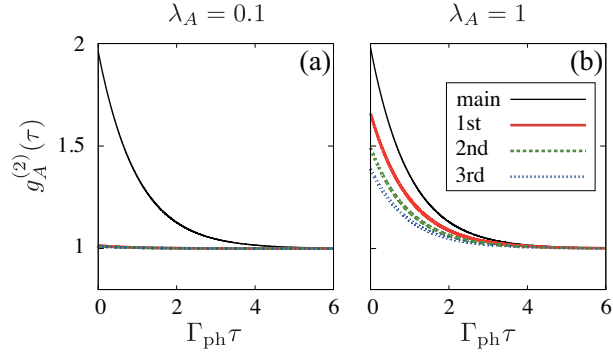}
    \caption{
    Autocorrelation function of $A$-phonons,
    $g^{(2)}_A (\tau)$, at the current main peak ($\Delta = 0$) and
    subpeaks ($\Delta = \Delta_{\nu} \simeq \nu \hbar \omega_{\rm ph}$,
    $\nu=1,2,3$) in figure \ref{fig:Delta}(a).
    (a) $\lambda_A = 0.1$ or (b) 1, and $\lambda_S = 0$.
    The other parameters are the same as in figure \ref{fig:Delta}.
    Note that three lines for the current subpeaks
    are almost overlapped in panel (a).
    \label{fig:g2_tau}}
\end{center}
\end{figure}

To compare the two situations in detail,
we present the number distribution of $A$-phonons in figures
\ref{fig:distribution}(a) and (b) at the current main peak
and subpeaks. In the case of $\lambda_A = 0.1$,
a Poisson-like distribution emerges at the subpeaks,
whereas a Bose distribution with effective temperature $T^*$
is seen at the main peak.
$T^*$ is determined from the number of phonons
$\langle N_A \rangle$ in the stationary state as
$
    1/[\rme^{\hbar \omega_{\rm ph}/(k_{\rm B} T^*)} - 1]
    = \langle N_A \rangle
$.
When $\lambda_A = 1$, on the other hand,
the distribution shows an intermediate shape between
Poisson and Bose distributions at the subpeaks
and the Bose distribution at the main peak.

In figures \ref{fig:g2_tau}(a) and (b),
we plot the autocorrelation function $g^{(2)}_A (\tau)$
as a function of $\tau$.
In the case of $\lambda_A = 0.1$, $g^{(2)}_A (\tau) \simeq 1$,
regardless of the time delay $\tau$, which
supports the phonon lasing at the current subpeaks.
At the main peak,
$g^{(2)}_A (\tau) \simeq 1 + \rme^{- \Gamma_{\rm ph} \tau}$. This is
a character of thermal phonons with temperature $T^*$.
When $\lambda_A = 1$, we find an intermediate behavior,
$g^{(2)}_A (\tau) \simeq 1 + \delta_\nu \rme^{- \Gamma_{\rm ph} \tau}$
($0 < \delta_\nu < 1$), at the $\nu$th subpeak.
This indicates that the phonons are partly thermalized
by the Franck-Condon effect. For larger $\nu$, the distribution
is closer to the Poissonian with smaller $\delta_\nu$.

\subsection{Competition between phonon lasing and
Franck-Condon thermalization}

To elucidate the competition between the phonon lasing and
thermalization by the Franck-Condon effect, we analyze the
rate equation in equation (\ref{eq:rate}), focusing on the current peaks
in the large bias-voltage limit.
We introduce  polaron states
$|L(R), n \rangle_{eA}$ for an electron in dot $L$ $(R)$ and $n$ phonons
with lattice distortion:
\begin{eqnarray}
    |L, n \rangle_{eA} = |L \rangle_e
        \otimes \mathcal{T}_A |n \rangle_A,
\quad
    |R, n \rangle_{eA} = |R \rangle_e
        \otimes \mathcal{T}_A^{\dagger} |n \rangle_A,
\label{eq:polaronA}
\end{eqnarray}
where
\begin{eqnarray}
    \mathcal{T}_A = \rme^{- \lambda_A (a_A^\dagger - a_A)}
\end{eqnarray}
and its Hermitian conjugate $\mathcal{T}_A^\dagger$
describe the shift of equilibrium position of the lattice
when an electron stays in dot $L$ and $R$, respectively.
Note that the lattice distortion produces $\lambda_A^2$ extra phonons:
$_{eA} \langle \alpha, n| N_A |\alpha, n \rangle_{eA} = n + \lambda_A^2$.
When $\Delta = \Delta_{\nu}$ ($\simeq \nu \hbar \omega_{\rm ph}$),
the eigenstates of Hamiltonian $H$ are given by the zero-electron states
$|0, n \rangle_{eA} = |0 \rangle_e \otimes |n \rangle_A$,
bonding and anti-bonding states
between the polarons,
\begin{eqnarray}
    |\pm, n \rangle_{eA} = \frac{1}{\sqrt2}
        (|L, n \rangle_{eA} \pm |R, n + \nu \rangle_{eA})
\end{eqnarray}
$(n = 0, 1, 2, \ldots)$, and polarons localized in dot $R$,
$|R, n \rangle$ ($n = 0, 1, 2, \ldots, \nu - 1$).
This is a good approximation
provided that $V_{\rm C} \ll \hbar \omega_{\rm ph}$.
The rate equations for these states are
\begin{eqnarray}
\fl \dot{P}_{0, n} = -\Gamma_L P_{0, n}
        + \sum_{m = 0}^{\infty} \frac{\Gamma_R}{2}
        |_A \langle n | \mathcal{T}_A^\dagger | m + \nu \rangle_A |^2
        P_{{\rm mol}, m}
        + \sum_{m = 0}^{\nu - 1} \Gamma_R
        |_A \langle n | \mathcal{T}_A^\dagger | m \rangle_A |^2 P_{R, m}
    \nonumber \\
        + \Gamma_{\rm ph} \left[ (n + 1) P_{0, n + 1} - n P_{0, n} \right],
        \label{eq:rate_0_A} \\
\fl \dot{P}_{{\rm mol}, n} = -\frac{\Gamma_R}{2} P_{{\rm mol}, n}
        + \sum_{m = 0}^{\infty} \Gamma_L
        |_A \langle n | \mathcal{T}_A | m \rangle_A |^2 P_{0, m}
    \nonumber \\
    + \Gamma_{\rm ph} \left[
        \left( n + 1 + \frac{\nu}{2} \right) P_{{\rm mol}, n + 1}
        - \left( n + \frac{\nu}{2} \right)
        P_{{\rm mol}, n} \right],
        \label{eq:rate_mol_A}
\end{eqnarray}
where $P_{{\rm mol}, n} = P_{+, n} + P_{-, n}$
($n = 0, 1, 2, \ldots$), and
\begin{eqnarray}
    \dot{P}_{R, n} &= - \Gamma_R P_{R, n}
        + \Gamma_{\rm ph} \bigl[ (n + 1) P_{R, n + 1} - n P_{R, n} \bigr],
        \label{eq:rate_R_A}
\end{eqnarray}
with $P_{R, \nu} = P_{{\rm mol}, 0}/2$ ($n = 0, 1, 2, \ldots, \nu - 1$).
As shown in \ref{app:A-phonon},
these equations yield the current $I$ and
electron number in the DQD,
$\langle n_e \rangle = \langle n_L + n_R \rangle$,
in terms of the number of polarons localized in dot $R$,
$\langle \tilde n_R \rangle = \sum_{n = 0}^{\nu - 1} P_{R, n}$, as
\begin{eqnarray}
    I = e \Gamma_R \frac{ 1 + \langle \tilde n_R \rangle }{2 + \gamma},
        \qquad
    \langle n_e \rangle = \frac{2 - \gamma \langle \tilde n_R \rangle}
                            {2 + \gamma},
        \label{eq:I_A}
\end{eqnarray}
with $\gamma = \Gamma_R / \Gamma_L$.
The number of $A$-phonons is given by
\begin{eqnarray}
    \langle N_A \rangle = (\nu + 2 \lambda_A^2) \frac{I}{e \Gamma_{\rm ph}}
        + \lambda_A^2 \langle n_e \rangle.
        \label{eq:N_A}
\end{eqnarray}
The first two terms in equation (\ref{eq:N_A})
indicate the emission of $\nu$ phonons by the
phonon-assisted tunneling (from dot $L$ to dot $R$)
and creation of $2 \lambda_A^2$ phonons by the lattice distortion
(with two tunnelings between the DQD and leads)
per transfer of a single electron  through the DQD.
The last term describes the average number of polarons
$\langle n_e \rangle$ in the stationary state.

When $\Gamma_{L,R} \gg \Gamma_{\rm ph}$, we obtain
\begin{eqnarray}
I = I_0 + \mathcal{O}(\Gamma_{\rm ph}/\Gamma_{L,R}),
        \label{eq:I_A2}
\end{eqnarray}
where $I_0 = e \Gamma_R / (2 + \gamma)$
is the current at the main peak in the absence of
electron-phonon interaction, and
\begin{eqnarray}
    g^{(2)}_A (0) = \frac{\nu + 4 \lambda_A^2}{\nu + 2 \lambda_A^2}
        + \mathcal{O} (\Gamma_{\rm ph}/\Gamma_{L,R}).
        \label{eq:g2_A}
\end{eqnarray}
These explain the numerical results in figure \ref{fig:Delta}
at the current subpeaks.
The formula in equation (\ref{eq:g2_A})
indicates $g^{(2)}_A (0) \simeq 1$ (phonon lasing)
for $\lambda_A^2 \ll \nu$ and
$g^{(2)}_A (0) \simeq 2$ (thermalized phonons by the lattice
distortion) for $\lambda_A^2 \gg \nu$. In the latter case,
the phonons follow the Bose distribution with $T^*$
to deduce $\langle N_A \rangle$ in equation (\ref{eq:N_A}).

We comment on the peak width of the electric current in
figure \ref{fig:Delta}(a).
The electron transfer around the $\nu$th current peak
is dominated by the tunneling between polaron states
$|L, n \rangle_{eA}$ and $|R, n + \nu \rangle_{eA}$
with $n \simeq \langle N_A \rangle$.
Thus the peak width is
determined by the effective tunnel coupling
\begin{eqnarray}
    W_\nu &= |_{eA} \langle R, n + \nu| H_e | L, n \rangle_{eA}|
        _{n \simeq \langle N_A \rangle}
        \nonumber \\
    & = \left| \frac{n!}{(n + \nu)!} (-2 \lambda_A)^\nu \rme^{-2 \lambda_A^2}
        L^\nu_n (4 \lambda_A^2) V_{\rm C} \right|
        _{n \simeq \langle N_A \rangle}
        \label{eq:W}
\end{eqnarray}
$(\nu = 0, 1, 2, \ldots)$, where $L^\nu_n (x)$ is
the Laguerre polynomial\footnote{
    In the absence of electron-phonon interaction,
    $\Delta$-dependence of the current shows a peak at $\Delta=0$. The peak
    width is given by the tunnel coupling $V_{\rm C}$ between the quantum
    dots [see equation (\ref{eq:I_S})]. In the presence of electron-phonon
    interaction, $V_{\rm C}$ is replaced by $W_\nu$ for the tunnel coupling
    between the polarons at the $\nu$th subpeak.
}.
The factor of $\rme^{-2 \lambda_A^2}$ in equation (\ref{eq:W}) stems
from the electron localization by dressing the phonons in forming
the polarons. This explains the narrower subpeaks in the case of
$\lambda_A=1$ than that of $\lambda_A=0.1$.

When $\lambda_A^2 \ll 1$, equation (\ref{eq:W}) yields
\begin{eqnarray}
    W_\nu \simeq \sqrt{\langle N_A \rangle + 1} \lambda_A^\nu V_{\rm C}.
        \label{eq:W_weak}
\end{eqnarray}
This is in quantitative accordance with the peak widths in the case of
$\lambda_A=0.1$ [solid line in figure \ref{fig:Delta}(a)].

\subsection{Franck-Condon blockade
    \label{subsec:FCB_A}}

\begin{figure*}
\begin{center}
    \includegraphics[width=12cm]{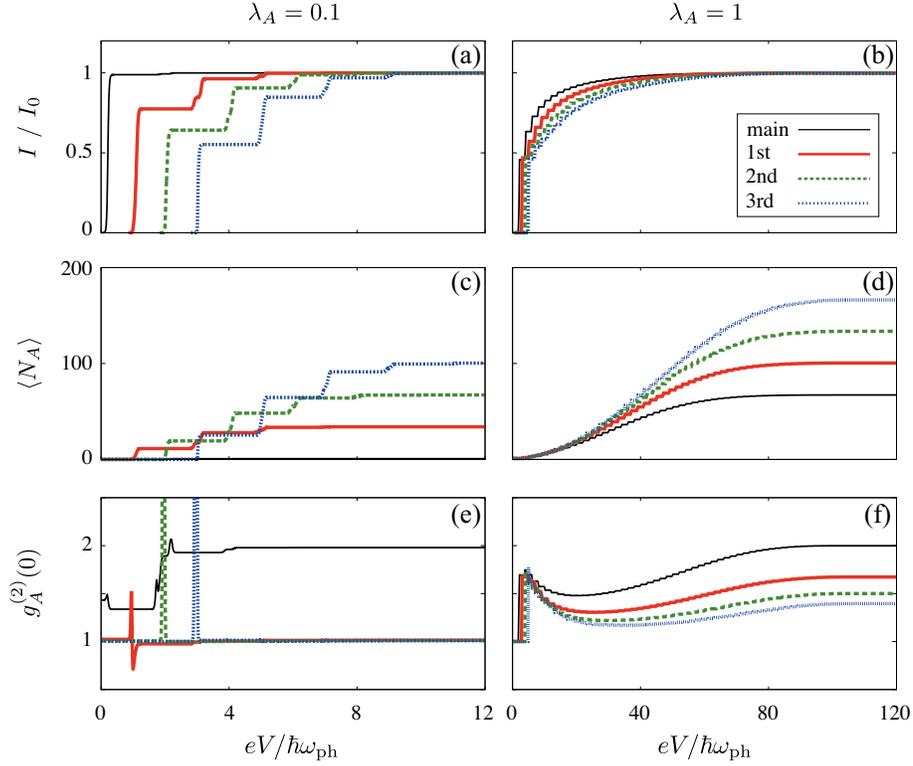}
    \caption{
    (a), (b) Electric current,
    (c), (d) $A$-phonon number $\langle N_A \rangle$,
    and (e), (f) its autocorrelation function $g^{(2)}_A (0)$,
    as a function of bias voltage $V$,
    at the current main peak ($\Delta=0$) and
    subpeaks ($\Delta=\Delta_{\nu} \simeq \nu \hbar \omega_{\rm ph}$,
    $\nu=1,2,3$) in figure \ref{fig:Delta}(a).
    $\lambda_A = 0.1$ in panels (a), (c), and (e), whereas
    $\lambda_A = 1$ in panels (b), (d), and (f).
    $\lambda_S = 0$ and $T = 0.01~\hbar \omega_{\rm ph}$.
    The other parameters are the same as in figure \ref{fig:Delta}.
    \label{fig:V_A}}
\end{center}
\end{figure*}

So far we have considered the large bias-voltage limit.
In this subsection, we examine the case of finite bias voltages
to elucidate the Franck-Condon blockade \cite{Sapmaz2006}
in our system.
Figures \ref{fig:V_A}(a) and (b) show the electric current
as a function of bias voltage $V$ when the level spacing $\Delta$
is tuned to the main and subpeaks in figure \ref{fig:Delta}(a).
The electron-phonon coupling is (a) $\lambda_A = 0.1$ and
(b) 1.
At the main peak ($\Delta = 0$) in case (a), the current
is almost identical to $I_0$ in the large bias-voltage limit when
$\mu_L = eV / 2$
exceeds the interdot tunnel coupling $V_{\rm C}$.
(The current vanishes when $eV / 2 \lesssim V_{\rm C}$,
reflecting the formation of
bonding and antibonding orbitals at energy level $\pm V_{\rm C}$,
from two orbitals at $\varepsilon_L=\varepsilon_R=0$ in the DQD.)
The influence of electron-phonon coupling is hardly observable.
At the subpeaks
($\Delta = \Delta_{\nu} \simeq \nu \hbar \omega_{\rm ph}$,
$\nu=1,2,3$) in case (a),
on the other hand, the current is suppressed at small $V$ and it
increases stepwise to the value in the large $V$ limit. This is due
to the electron-phonon coupling, as explained below.
The current suppression is much more prominent in case (b) with larger
$\lambda_A$. We observe the suppression even at the main peak in this case.

The reason for the current suppression is as follows.
When an electron tunnels between the DQD and leads,
the equilibrium position of the lattice is suddenly changed
to form the polaron, $|L, n \rangle_{eA}$ or $|R, n \rangle_{eA}$,
in equation (\ref{eq:polaronA}).
While all the phonon states participate in the polaron formation in the
large bias-voltage limit, the phonon states are limited under finite
bias voltages due to the energy conservation. This weakens the tunnel
coupling between the DQD and leads and also between the quantum dots,
which is known as the Franck-Condon blockade.
In figures \ref{fig:V_A}(a) and (b), the current
increases stepwise as $\mu_L = eV / 2$ increases by
$\hbar \omega_{\rm ph}$ because
higher-energy states become accessible (Franck-Condon steps)
and converges to $I = I_0$ in the large bias-voltage limit.
The larger voltage is required to lift off the Franck-Condon blockade
for larger $\lambda_A$
    \cite{Koch2005}.

Figures \ref{fig:V_A}(c)--(f) show the phonon number and autocorrelation
function $g^{(2)}_A (0)$ as a function of $V$.
The phonon number shows the Franck-Condon steps in both cases of
(c) $\lambda_A = 0.1$ and (d) $1$. The autocorrelation function,
on the other hand, is qualitatively different for the two cases.
In figure \ref{fig:V_A}(e) with $\lambda_A = 0.1$,
$g^{(2)}_A (0) \simeq 1$ even at the first Franck-Condon step
except for anomalous behavior around the beginning of the step.
This indicates that the phonon lasing is robust against the
current suppression by the Franck-Condon blockade and hence
it is observable under finite bias.
In figure \ref{fig:V_A}(f) with $\lambda_A = 1$,
$g^{(2)}_A (0)$ changes slowly with $V$, reflecting
$V$-dependence of the thermalization due to the Franck-Condon effect.

\begin{figure}
\begin{center}
    \includegraphics[width=10cm]{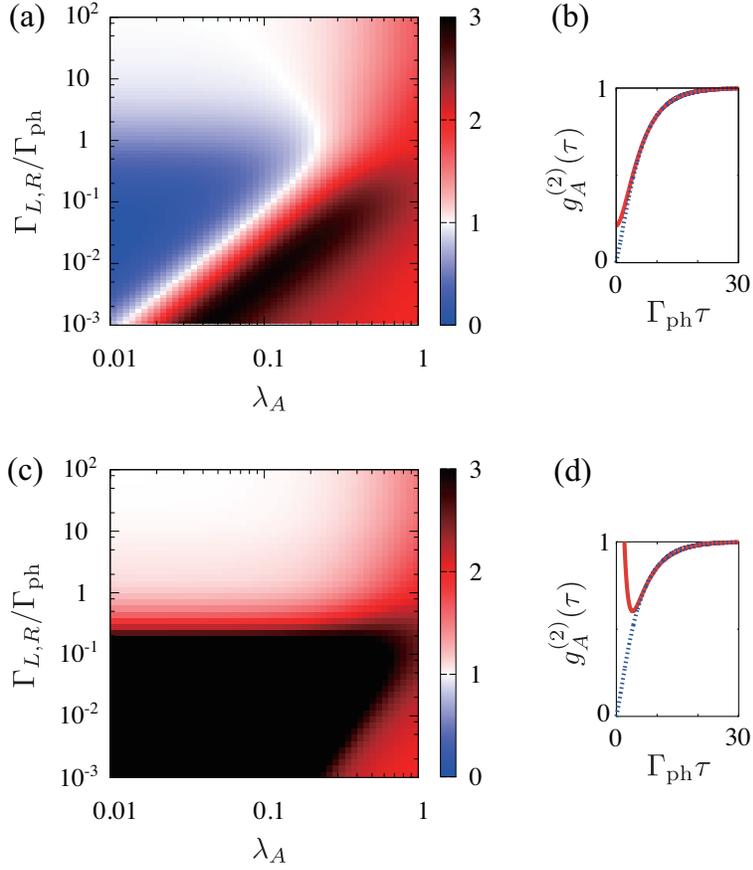}
    \caption{
    (a), (b) Color-scale plot of autocorrelation function of $A$-phonons,
    $g^{(2)}_A (0)$, at the current subpeaks
    ($\Delta = \Delta_{\nu} \simeq \nu \hbar \omega_{\rm ph}$, $\nu=1,2$)
    in the large bias-voltage limit,
    in a plane of electron-phonon coupling $\lambda_A$ and
    $\Gamma_{L, R}/\Gamma_{\rm ph}$.
    $\Gamma_L = \Gamma_R \equiv \Gamma_{L, R}$,
    $\lambda_S = 0$, and $V_{\rm C} = 0.1~\hbar \omega_{\rm ph}$.
    (c), (d) $g^{(2)}_A (\tau)$ at $\lambda_A = 0.05$ and
    $\Gamma_{L,R} = 0.1~\Gamma_{\rm ph}$, as a function
    of $\tau$ (solid line). The autocorrelation function of
    electric current, $g^{(2)}_{\rm current} (\tau)$, is also shown by
    dotted line.
    Panels (a) and (b) are indicated for the first current subpeak ($\nu=1$),
    whereas panels (c) and (d) are for the second current subpeak
    ($\nu=2$).
    \label{fig:antibunching}}
\end{center}
\end{figure}

\subsection{Phonon antibunching
    \label{subsec:antibunching}}

In subsections \ref{subsec:lasing} to \ref{subsec:FCB_A},
we have restricted ourselves to the case
of $\Gamma_{L, R} \gg \Gamma_{\rm ph}$ to examine the phonon lasing.
If the tunnel coupling is tuned to be
$\Gamma_{L, R} \lesssim \Gamma_{\rm ph}$,
we observe another phenomenon, antibunching of $A$-phonons
    \cite{Lambert2008}.
Figure \ref{fig:antibunching}(a) presents a color-scale plot of
$g^{(2)}_A(0)$ in the $\lambda_A$--$(\Gamma_{L, R}/\Gamma_{\rm ph})$ plane
when $\Delta$ is tuned to be at the first current subpeak
($\Delta = \Delta_1 \simeq \hbar \omega_{\rm ph}$).
We assume that $\Gamma_L = \Gamma_R \equiv \Gamma_{L, R}$, $\lambda_S=0$,
and large limit of bias voltage.
At $\lambda_A=0.05$ and $\Gamma_{L, R}/\Gamma_{\rm ph}=0.1$, for example,
$g^{(2)}_A (0) \ll 1$, representing a strong antibunching of phonons.
This is because the phonon emission is regularized by the
electron transport through the DQD.
In figure \ref{fig:antibunching}(b),
we plot the autocorrelation function of the electric current
\begin{eqnarray}
    g^{(2)}_{\rm current} (\tau) =
    \frac{\langle :n_R(0) n_R(\tau): \rangle}{\langle n_R \rangle^2},
\end{eqnarray}
where $n_R$ is the electron number in dot $R$.
It fulfills $g^{(2)}_{\rm current} (0)=0$, indicating the antibunching of
electron transport, since dot $R$ is empty just after
the electron tunnels out
    \cite{Emary2012}.
Remarkably, $g^{(2)}_A(\tau)$ almost coincides
with $g^{(2)}_{\rm current} (\tau)$. When $\Gamma_{L, R} \ll \Gamma_{\rm ph}$,
the emitted phonon escapes from the natural cavity soon after the electron
tunneling between the quantum dots. Thus the stimulated emission for
the lasing does not take place.

At strong couplings of $\lambda_A \gtrsim 1$,
neither phonon antibunching nor phonon lasing can be
observed because of an effective phonon thermalization
due to the Franck-Condon effect. More than one phonon is created
by the polaron formation, which spoils the regularized phonon
emission by single electron tunneling and results in the phonon
bunching.

Even with small $\lambda_A$, bunched phonons are emitted if 
$\Gamma_{L, R}/\Gamma_{\rm ph}$ is too small. Then the number of phonons
created by the tunneling is exceeded by that
accompanied by the polaron staying in dot $R$ [the first two terms are
much smaller than the last term in equation (\ref{eq:N_A})],
as discussed in \ref{app:large-decay-phonon}. The analytical
expression of $g^{(2)}_A(0)$ is also given for
$\Gamma_{L, R} \ll \Gamma_{\rm ph}$ in the appendix.

Figure \ref{fig:antibunching}(c) shows a color-scale plot of
$g^{(2)}_A (0)$ when $\Delta$ is tuned to be at the second current
subpeak ($\Delta = \Delta_2 \simeq 2 \hbar \omega_{\rm ph}$).
The antibunching does not occur even when
$\Gamma_{L, R} \ll \Gamma_{\rm ph}$ because two phonons are emitted
simultaneously by the electron tunneling, which are bunched to
each other.

\section{Franck-Condon effect of $S$-phonons
    \label{sec:S-phonon}}

In this section, we examine $S$-phonons
and disregard $A$-phonons with $\lambda_A = 0$.

\subsection{Franck-Condon thermalization}

We begin with the large bias-voltage limit.
The electric current has a single-peaked structure
as a function of $\Delta$ [Lorentzian with center at $\Delta=0$
and width of $V_{\rm C}\sqrt{2+\gamma}$, as will be seen in equation
(\ref{eq:I_S})].
We do not observe subpeaks at $\Delta \simeq \nu \hbar \omega_{\rm ph}$
since $S$-phonons are not relevant to the phonon-assisted
tunneling between the quantum dots because they couple to
the total number of electron, $n_L + n_R$ in the DQD.
The polaron states involving $S$-phonons are given by
\begin{eqnarray}
    |L, n \rangle_{eS} = |L \rangle_e \otimes \mathcal{T}_S |n \rangle_S,
\quad
    |R, n \rangle_{eS} = |R \rangle_e \otimes \mathcal{T}_S |n \rangle_S,
\label{eq:polaronS}
\end{eqnarray}
for an electron in dot $L$ or $R$, with $n$ phonons, where
the lattice distortion
\begin{eqnarray}
    \mathcal{T}_S = \rme^{- \lambda_S (a_S^\dagger - a_S)}
\end{eqnarray}
is common for $|L, n \rangle_{eS}$ and $|R, n \rangle_{eS}$.
$S$-phonons do not show the phonon lasing nor antibunching.

We derive the rate equation for
arbitrary level spacing $\Delta$ in \ref{app:S-phonon}.
By tracing out $S$-phonon degrees of freedom,
we obtain the reduced rate equation for electrons,
which is the same as that in the absence of electron-phonon coupling.
We obtain the electric current and electron number in the DQD,
\begin{eqnarray}
    I &= \frac{e \Gamma_R}{(\Delta / V_{\rm C})^2 + 2 + \gamma},
    \qquad
    \langle n_e \rangle &= \frac{(\Delta / V_{\rm C})^2 + 2}
        {(\Delta / V_{\rm C})^2 + 2 + \gamma}.
        \label{eq:I_S}
\end{eqnarray}
The number of $S$-phonons is given by
\begin{eqnarray}
    \langle N_S \rangle = 2 \lambda_S^2 \frac{I}{e \Gamma_{\rm ph}}
        + \lambda_S^2 \langle n_e \rangle.
        \label{eq:N_S}
\end{eqnarray}
The first term in equation (\ref{eq:N_S}) indicates
the creation of $2 \lambda_S^2$ phonons by
the lattice distortion with two tunnelings between the DQD and leads
per a single electron transfer through the DQD.
The second term describes the average number of polarons.
In contrast to equation (\ref{eq:N_A}) for $A$-phonons,
$S$-phonons are not created by the interdot tunneling.

We also examine the autocorrelation function
\begin{eqnarray}
    g^{(2)}_S (\tau) = \frac{\langle : N_S(0) N_S(\tau) : \rangle}
        {\langle N_S \rangle^2}.
\end{eqnarray}
$g^{(2)}_S (0)$ is independent of $\lambda_S$,
for arbitrary $\Delta$.
When $\Gamma_{L, R} / \Gamma_{\rm ph} \gg 1$, we find
\begin{eqnarray}
    g^{(2)}_S (0) = 2 + \mathcal{O}(\Gamma_{\rm ph} / \Gamma_{L, R}),
    \label{eq:g2_S}
\end{eqnarray}
which indicates the thermalization induced by
the Franck-Condon effect.

\begin{figure}
\begin{center}
    \includegraphics[width=12cm]{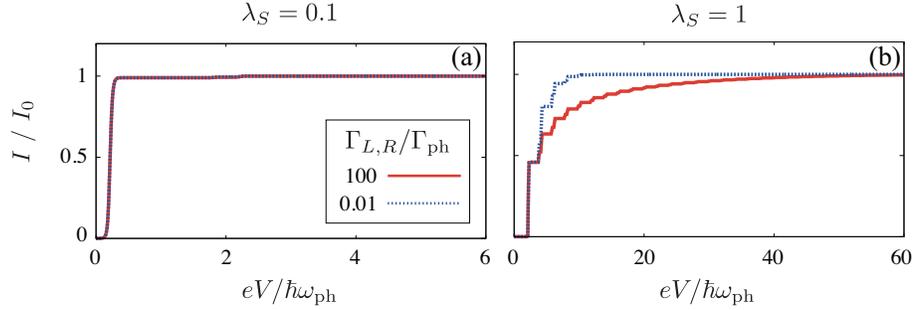}
    \caption{
    Electric current through the DQD at $\Delta = 0$,
    as a function of bias voltage $V$.
    The dimensionless electron-phonon coupling constants are
    (a) $\lambda_S = 0.1$ or (b) $1$, and $\lambda_A = 0$.
    $\Gamma_L = \Gamma_R \equiv \Gamma_{L,R}$,
    $V_{\rm C} = 0.1~\hbar \omega_{\rm ph}$,
    and $T = 0.01~\hbar \omega_{\rm ph}$.
    Note that two lines are almost overlapped in panel (a).
    \label{fig:V_S}}
\end{center}
\end{figure}

\subsection{Franck-Condon blockade
    \label{subsec:FCB_S}}

Next, we examine the Franck-Condon blockade under finite bias
voltages. Figures \ref{fig:V_S}(a) and (b) show
the current as a function of $V$ when $\Delta = 0$.
The dimensionless coupling constant is (a) $\lambda_S = 0.1$ and (b) 1.
While the Franck-Condon blockade suppresses the current
under small bias voltages in the case of $\lambda_S = 1$,
the current suppression is negligible in the case of $\lambda_S = 0.1$.
In the former case, a larger voltage is needed to lift off the
Franck-Condon blockade for larger $\Gamma_{L, R} / \Gamma_{\rm ph}$.

When $\Gamma_{L, R} / \Gamma_{\rm ph}=100$,
the $V$ dependence of the current is almost the same as in
figure \ref{fig:V_A} for $A$-phonons with $\Delta=0$.
The phonon number and its autocorrelation function
also change with the bias voltage $V$ in a similar manner to
those at the current main peak for $A$-phonons.

\section{Coupling with both phonon modes
    \label{sec:AS-phonon}}

Now we consider both $A$- and $S$-phonons.
Here, we examine a DQD fabricated in the semiconductor substrate
where an electron is weakly coupled to both phonons;
$\lambda_S$, $\lambda_A \lesssim 0.1$.

First, we analytically derive that the coupling to $S$-phonons does not
influence the calculated results in section \ref{sec:A-phonon}
for the electron-$A$-phonon system in the large bias-voltage limit.
Consider the current main peak ($\Delta =0$)
and subpeaks ($\Delta=\Delta_{\nu} \simeq \nu \hbar \omega_{\rm ph}$),
assuming that $V_{\rm C} \ll \hbar \omega_{\rm ph}$.
The eigenstates of $H$ are given by the zero-electron states
$|0, n; n' \rangle_{eA;S} = |0, n \rangle_{eA} \otimes |n' \rangle_S$,
bonding and anti-bonding states between the polarons,
$
    |\pm, n; n' \rangle_{eA;S} = |\pm, n \rangle_{eA}
        \otimes \mathcal{T}_S |n' \rangle_S
$
$(n, n' = 0, 1, 2, \ldots)$, and polarons localized in dot $R$,
$
    |R, n; n' \rangle_{eA;S} = |R, n \rangle_{eA}
        \otimes \mathcal{T}_S |n' \rangle_S
$
$(n = 0, 1, 2, \ldots, \nu; n' = 0, 1, 2, \ldots)$.
The rate equations for these states yield
\begin{eqnarray}
\fl \dot{P}_{0, n; n'} = -\Gamma_L P_{0, n; n'}
        + \sum_{m, m' = 0}^{\infty} \frac{\Gamma_R}{2}
        |_A \langle n | \mathcal{T}_A^\dagger | m + \nu \rangle_A |^2
        |_S \langle n' | \mathcal{T}_S^\dagger | m' \rangle_S |^2
        P_{{\rm mol}, m; m'}
        \nonumber \\
    + \sum_{m = 0}^{\nu - 1} \sum_{m' = 0}^\infty \Gamma_R
        |_A \langle n | \mathcal{T}_A^\dagger | m \rangle_A |^2
        |_S \langle n' | \mathcal{T}_S^\dagger | m' \rangle_S |^2
        P_{R, m; m'}
        \nonumber \\
    + \Gamma_{\rm ph} \left[ (n + 1) P_{0, n + 1; n'}
        + (n' + 1) P_{0, n; n' + 1} - (n + n') P_{0, n; n'} \right],
        \label{eq:rate_0_AS} \\
\fl \dot{P}_{{\rm mol}, n; n'} = -\frac{\Gamma_R}{2} P_{{\rm mol}, n; n'}
        + \sum_{m,m' = 0}^{\infty} \Gamma_L
        |_A \langle n | \mathcal{T}_A | m \rangle_A |^2
        |_S \langle n' | \mathcal{T}_S | m' \rangle_S |^2
        P_{0, m; m'}
        \nonumber \\
\fl \qquad + \Gamma_{\rm ph} \left[
        \left( n + 1 + \frac{\nu}{2} \right) P_{{\rm mol}, n + 1; n'}
        + (n' + 1) P_{{\rm mol}, n; n'+1}
        - \left( n + n' + \frac{\nu}{2} \right) P_{{\rm mol}, n; n'} \right],
        \label{eq:rate_mol_AS}
\end{eqnarray}
where $P_{{\rm mol}, n; n'} = P_{+, n; n'} + P_{-, n; n'}$
($n, n' = 0, 1, 2, \ldots$), and
\begin{eqnarray}
\fl \dot{P}_{R, n; n'} &= - \Gamma_R P_{R, n; n'}
        + \Gamma_{\rm ph} \bigl[ (n + 1) P_{R, n + 1; n'}
        + (n' + 1) P_{R, n; n' + 1} - (n + n') P_{R, n; n'} \bigr],
        \label{eq:rate_R_AS}
\end{eqnarray}
with $P_{R, \nu; n'} = P_{{\rm mol}, 0; n'} / 2$
$(n = 0, 1, 2, \ldots, \nu - 1; n' = 0, 1, 2, \ldots)$.
We trace out $S$-phonon degrees of freedom by summing up
both sides of equations (\ref{eq:rate_0_AS})--(\ref{eq:rate_R_AS})
over $n'$.
We then obtain the reduced rate equations for the electron-$A$-phonon
system, which are just the same as equations
(\ref{eq:rate_0_A})--(\ref{eq:rate_R_A})
with $P_{0, n} = \sum_{n'=0}^\infty P_{0, n; n'}$,
$P_{{\rm mol}, n} = \sum_{n'} P_{{\rm mol}, n; n'}$,
and $P_{R, n} = \sum_{n'} P_{R, n; n'}$.
This fact indicates that $S$-phonons do not affect
the dynamics of the electron-$A$-phonon system if the bias voltage is
sufficiently large.

\begin{figure}
\begin{center}
    \includegraphics[width=6.5cm]{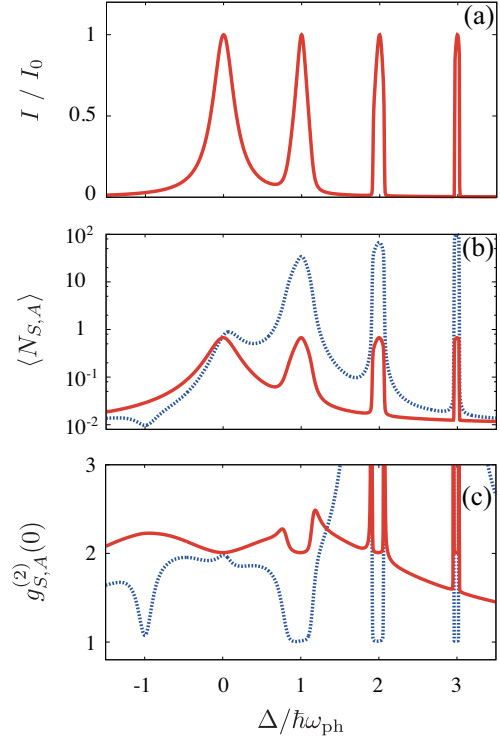}
    \caption{
    (a) Electric current through the DQD,
    (b) phonon numbers $\langle N_{S,A} \rangle$,
    and (c) the autocorrelation function
    $g^{(2)}_{S,A} (0)$ in the large bias-voltage limit,
    as a function of level spacing $\Delta$ in the
    DQD. In panels (b) and (c), data for $S$-phonons
    ($A$-phonons) are indicated by solid (dotted) lines.
    The dimensionless electron-phonon coupling constants are
    $\lambda_A = \lambda_S = 0.1$. In panel (a),
    $I_0 = e \Gamma_R / (2 + \Gamma_R / \Gamma_L)$ is the
    current at $\Delta=0$
    in the absence of electron-phonon coupling.
    $\Gamma_L = \Gamma_R = 100~\Gamma_{\rm ph}$
    and $V_{\rm C} = 0.1~\hbar \omega_{\rm ph}$.
    \label{fig:Delta_AS}}
\end{center}
\end{figure}

In figure \ref{fig:Delta_AS}, we plot (a) the electric current, (b)
$A$- and $S$-phonon numbers, and (c) their autocorrelation function,
as a function of $\Delta$, in the case of $\lambda_A = \lambda_S = 0.1$
and $\Gamma_{L, R} \gg \Gamma_{\rm ph}$.
The current, phonon number, and autocorrelation function for
$A$-phonons are the same as in figure \ref{fig:Delta} with
$\lambda_A=0.1$ (solid line) where $S$-phonons are disregarded,
in accordance with the above-mentioned consideration.
An increase in $S$-phonon number $\langle N_S \rangle$
is induced by the current via the Franck-Condon effect.
It is explained by equation (\ref{eq:N_S}) using the current $I$
and electron number $\langle n_e \rangle$.
$g^{(2)}_S (0) \simeq 2$ at the current peaks, indicating
the thermalization of $S$-phonons.

When the bias voltage is finite, $S$-phonon degrees of freedom cannot
be traced out in the rate equation. Therefore $S$-phonons can
influence the current and distribution of $A$-phonons.
However, the influence is very small, provided that
$\lambda_S \sim \lambda_A \lesssim 0.1$, because the current suppression
by the Franck-Condon blockade with $S$-phonons is negligible,
as shown in figure \ref{fig:V_S}(a).

\section{Discussion
    \label{sec:discussion}}

In the present work,
we consider single energy levels in quantum dots, $\varepsilon_L$
and $\varepsilon_R$ ($\Delta=\varepsilon_L-\varepsilon_R$).
We take into account the optical phonons but do not the acoustic phonons.
When $\Delta \sim \hbar \omega_{\rm ph}$ ($=36$ meV in GaAs),
there are several energy levels in dot $R$ between $\varepsilon_R$ and
$\varepsilon_L$. Thus some transport processes should exist in which
an electron tunnels from $\varepsilon_L$ to excited levels in dot $R$
with emitting LA phonons.
These LA-phonon-assisted tunneling processes, however, can be
neglected around the current subpeaks at $\Delta = \Delta_{\nu}$
($\simeq \nu \hbar \omega_{\rm ph}$) in figure \ref{fig:Delta}(a), where
the LO-phonon-assisted tunneling processes are dominant.
The reason is as follows.
In quantum dots of radius $\mathcal{R}$,
electrons are coupled to acoustic phonons with small wavenumbers of
$|{\bi q}| \lesssim 1/\mathcal{R}$ only\footnote{
    The electron-LA-phonon coupling is described by the
    piezoelectric or deformation potential.
    In both cases, the coupling constant involves the integral,
    $
        \int \rmd {\bi r} |\psi_\alpha ({\bi r})|^2
        \rme^{\rmi {\bi q} \cdot {\bi r}}
    $,
    for an electron in dot $\alpha$ and LA phonon with wavenumber $\bi q$,
    as in the case of electron-LO-phonon coupling in equation
    (\ref{eq:M}).
}.
When $\mathcal{R} < 100~{\rm nm}$, the energy of such LA phonons is
comparable to or smaller than the level spacing in the quantum dot.
Therefore, the number of relevant excited levels in dot $R$
is zero or unity.
Besides, the coupling to LA phonons is much weaker than that to LO
phonons because of large density of states in the latter.
Indeed the LO-phonon-assisted transport was clearly observed
for level spacings $\Delta$ tuned to $\hbar \omega_{\rm ph}$ and
$2\hbar \omega_{\rm ph}$ in recent experiments
    \cite{Amaha2012}.

Next, we discuss the validity of the BMS approximation.
The BMS approximation is based on the assumption
that the typical time scale
described by the Hamiltonian, $H$ in equation (\ref{eq:H_eff}),
is much larger than $1 / \Gamma_{L, R}$ and $1 / \Gamma_{\rm ph}$
    \cite{Breuer}.
At the $\nu$th current subpeak in semiconductor-based DQDs,
the typical time scale is estimated to be
$\hbar / W_\nu$ in equation (\ref{eq:W_weak}). Therefore,
our results on the phonon lasing are justified when
$V_{\rm C} \gg \hbar (\Gamma_{L, R} \Gamma_{\rm ph})^{1/2} / \lambda_A^\nu$,
and on the phonon antibunching at the first subpeak
when $V_{\rm C} \gg \hbar \Gamma_{\rm ph} / \lambda_A$.
We believe that our results are asymptotically applicable for
smaller $V_{\rm C}$.

Finally, we address possible experimental realizations to
observe LO phonon lasing and antibunching in semiconductor-based DQDs.
In GaAs, an LO phonon around the $\Gamma$ point decays into
an LO phonon and a TA phonon around the L point,
which are not coupled to the DQD.
These daughter phonons can be detected by the transport through another
DQD fabricated nearby
    \cite{Gasser2009,Harbusch2010}.
Alternatively, the modulation of the dielectric constant by the
phonons could be observed by near-field spectroscopy
    \cite{Cunningham2008}.
With a decay rate $\Gamma_{\rm ph} \sim 0.1~{\rm THz}$ in GaAs
    \cite{Vallee1994},
however, the lasing condition $\Gamma_{L, R} \gg \Gamma_{\rm ph}$
might be hard to realize. Other materials with longer lifetime of
optical phonons, such as ZnO
    \cite{Aku-Leh2005},
may be preferable to observe the phonon lasing.

\section{Conclusions
    \label{sec:conclusion}}

We have proposed the optical phonon lasing in
a semiconductor-based DQD under a finite bias voltage,
without any requirement of an additional cavity or resonator.
First, we have shown that only two phonon modes
($S$- and $A$-phonons) are coupled to the DQD, which act as a cavity
because of the flat dispersion relation of the optical phonons.
The electric transport is accompanied by $A$-phonon emission when
the energy level spacing in the DQD is tuned to the phonon energy.
This results in the phonon lasing when the tunneling rate
$\Gamma_{L, R}$ between the DQD and leads is much larger than
phonon decay rate $\Gamma_{\rm ph}$. We also find the antibunching of
$A$-phonons in the same system when
$\Gamma_{L, R} \lesssim \Gamma_{\rm ph}$.
Both effects are robust against the finite coupling to $S$-phonons.

For a DQD fabricated in a carbon nanotube, we have shown that
the lasing and antibunching are spoilt by bunched phonons
created by the Franck-Condon effect, due to the strong electron-phonon
coupling. The coupling also brings about the suppression of the
electric current, called Franck-Condon blockade, under finite bias
voltages.

Our fundamental research of LO phonon statistics is
also applicable to a freestanding semiconductor membrane
as a phonon cavity
    \cite{Weig2004,Ogi2010},
in which a resonating mode plays a role of LO phonons in our theory.
Since our theory gives conditions for lasing or
antibunching on the electron-phonon coupling and
tunneling rate, it would be useful to design
a cavity to generate various quantum states.
This would lead to new development of nanoelectromechanical systems (NEMS).

\ack

The authors acknowledge fruitful discussion with
K.\ Ono, S.\ Amaha,
Y.\ Kayanuma, K.\ Saito,
C.\ P\"oltl, T.\ Yokoyama, and A.\ Yamada.
This work was partially supported by KAKENHI
(Nos.\ 23104724 and 24-6574),
Institutional Program for Young Researcher Oversea Visits
and International Training Program
from the Japan Society for the Promotion of Science,
Graduate School Doctoral Student Aid Program from Keio University,
and the German DFG via SFB 910 and project BR 1528/8-1.

\appendix

\section{Phonon mode function and coupling constant
    \label{app:mode_function}}

In this appendix, we derive phonon mode functions,
$\bi{u}_{S} (\bi{r})$ and $\bi{u}_{A} (\bi{r})$, shown in
figure \ref{fig:model}(b).
We also estimate the coupling constants
$\lambda_{S/A}$ in equation (\ref{eq:lambda}).

Using the optical phonon modes $a_{\bi{q}}$,
the lattice displacement at position $\bi{r}$ is given by
\begin{eqnarray}
    \bi{u} (\bi{r}) = -i \left( \frac{\hbar}{2 N \mu \omega_{\rm ph}}
         \right)^{1/2} \sum_{\bi{q}}
        \frac{\bi{q}}{q} \rme^{\rmi \bi{q} \cdot \bi{r}}
        (a_{\bi{q}} + a_{\bi{q}}^\dagger),
        \label{eq:app_u}
\end{eqnarray}
where $\mu$ is the reduced mass for a pair of Ga and As atoms in the
case of GaAs and $N$ is the number of pairs in the substrate.
The mode functions are defined as the coefficients of the $S$- and
$A$-phonons in the lattice displacement, i.e.,
\begin{eqnarray}
    \bi{u} (\bi{r}) =
        \bi{u}_S (\bi{r}) (a_S + a_S^\dagger)
      + \bi{u}_A (\bi{r}) (a_A + a_A^\dagger)
      + ({\rm other~modes}).
        \label{eq:app_u2}
\end{eqnarray}

From equations (\ref{eq:a_La_R}) and (\ref{eq:a_Sa_A}), $a_S$
and $a_A$ are expressed as
\begin{eqnarray}
    a_{S/A} =
    \frac{\sum_{\bi q} \left( M_{L, {\bi q}} \pm M_{R, {\bi q}} \right)
        a_{\bi q}}
    {\left( \sum_{\bi q} |M_{L, {\bi q}} \pm M_{R, {\bi q}}|^2 \right)^{1/2}}.
    \label{eq:app_a_Sa_A}
\end{eqnarray}
Here, we have used
\begin{eqnarray}
    \mathcal{S} = \frac{\sum_{\bi q} M_{L, {\bi q}} M_{R, -{\bi q}}}
    { \left( \sum_{\bi q} |M_{L, {\bi q}}|^2 \right)^{1/2}
      \left( \sum_{\bi q} |M_{R, {\bi q}}|^2 \right)^{1/2} }.
    \nonumber
\end{eqnarray}
From equation (\ref{eq:app_a_Sa_A}),
$a_{\bi q}$ is inversely expanded by $a_S$, $a_A$, and other modes:
\begin{eqnarray}
    \fl
    a_{\bi q} = \frac{(M_{L, -\bi{q}} + M_{R, -\bi{q}}) a_S}
    {\left( \sum_{\bi{q}'} |M_{L, {\bi{q}'}}
        + M_{R, {\bi{q}'}}|^2 \right)^{1/2}}
              + \frac{(M_{L, -\bi{q}} - M_{R, -\bi{q}}) a_A}
    {\left( \sum_{\bi{q}'} |M_{L, {\bi{q}'}}
        - M_{R, {\bi{q}'}}|^2 \right)^{1/2}}
              + \ldots,
\end{eqnarray}
By the substitution of this equation into (\ref{eq:app_u}), we obtain
\begin{eqnarray}
\fl
    \bi{u}_{S/A} (\bi{r}) &= -\rmi
        \left( \frac{\hbar}{2 N \mu \omega_{\rm ph}} \right)^{1/2}
        \frac{\sum_{\bi{q}} \frac{\bi{q}}{q} \rme^{\rmi \bi{q} \cdot \bi{r}}
              (M_{L, -{\bi q}} \pm M_{R, -{\bi q}})}
             {\left( \sum_{\bi q} |M_{L, {\bi q}}
              \pm M_{R, {\bi q}}|^2 \right)^{1/2}}
    \nonumber \\
\fl
    &= \left( \frac{\hbar \mathcal{V}}
       {8 \pi N \mu \omega_{\rm ph}} \right)^{1/2}
       \frac{1}{C_{S/A}}
          \int \rmd\bi{r}' \frac{
            [|\psi_L (\bi{r}')|^2 \pm |\psi_R (\bi{r}')|^2] (\bi{r} - \bi{r}')}
            {|\bi{r} - \bi{r}'|^3}
    \label{eq:u_Su_A}
\end{eqnarray}
with
\begin{eqnarray}
    C_{S/A}^2 &=
    \int \rmd \bi{r} \rmd \bi{r}'
    \frac{\left[ |\psi_L (\bi{r})|^2 \pm |\psi_R (\bi{r})|^2 \right]
    \left[ |\psi_L (\bi{r}')|^2 \pm |\psi_R (\bi{r}')|^2 \right]}
    {|\bi{r} - \bi{r}'|}.
    \label{eq:C_SC_A}
\end{eqnarray}
In the derivation of equations (\ref{eq:u_Su_A}) and (\ref{eq:C_SC_A}),
we have used
\begin{eqnarray}
    \int \frac{\rmd \bi{q}}{(2 \pi)^3}
        \frac{1}{q^2} \rme^{\rmi \bi{q} \cdot \bi{r}}
    = \frac{1}{4 \pi |\bi{r}|},
    \nonumber
\end{eqnarray}
and its gradient with respect to $\bi{r}$,
\begin{eqnarray}
    \rmi \int \frac{\rmd \bi{q}}{(2 \pi)^3}
        \frac{\bi{q}}{q^2} \rme^{\rmi \bi{q} \cdot \bi{r}}
    = - \frac{\bi{r}}{4 \pi |\bi{r}|^3}.
    \nonumber
\end{eqnarray}
In figure \ref{fig:model}(b), we evaluate the mode functions
in equation (\ref{eq:u_Su_A}),
assuming spherical Gaussian functions of radius $\mathcal{R}$
for the electron distribution in the quantum dots,
$|\psi_L (\bi{r})|^2$ and $|\psi_R (\bi{r})|^2$.

The coupling constants $\lambda_{S/A}$ in equation (\ref{eq:lambda})
are written as
\begin{eqnarray}
    \lambda_{S/A}^2 &= \frac{e^2}{32 \pi \hbar \omega_{\rm ph}}
    \left[ \frac{1}{\epsilon(\infty)} - \frac{1}{\epsilon(0)} \right]
C_{S/A}^2,
\end{eqnarray}
with $C_{S/A}$ in equation (\ref{eq:C_SC_A}).
Using the spherical Gaussian functions for $|\psi_L (\bi{r})|^2$ and
$|\psi_R (\bi{r})|^2$, we find
\begin{eqnarray}
    \lambda_{S/A}^2 & \simeq \frac{1}{\sqrt{32 \pi^3}}
    \left[ \frac{1}{\epsilon(\infty)} - \frac{1}{\epsilon(0)} \right]
    \left( 1 \pm \sqrt{\frac{\pi}{2}} \frac{\mathcal{R}}{d} \right)
    \frac{e^2}{\hbar \omega_{\rm ph} \mathcal{R}},
\end{eqnarray}
where $d = |\bi{r}_{LR}|$ is the distance between
centers of the two quantum dots.
This yields $\lambda_{S/A} = 0.01 \sim 0.1$ for
$\mathcal{R} = 100 \sim 10~{\rm nm}$ and $d \gtrsim 2 \mathcal{R}$,
in the case of GaAs.

\section{Effective Hamiltonian for double quantum dot in
carbon nanotube
    \label{app:CNT}}

In this appendix, we derive the effective Hamiltonian for a DQD
embedded in a suspended CNT.
An electron in the CNT is strongly coupled to the longitudinal
stretching modes (LSM) of phonons, known as vibrons,
by the deformation potential
    \cite{Mariani2009}.
We assume that quantum dots $L$ and $R$ are fabricated around
$x = x_L$ and $x_R$, respectively, in $0 < x < \ell$ along the CNT.
The phonon-related parts of the Hamiltonian are given by
\begin{eqnarray}
    \mathcal{H}_{\rm ph} &= \sum_{n = 1}^\infty
        \hbar \omega_n a_n^\dagger a_n, \\
    \mathcal{H}_{\rm ep} &= \sum_{\alpha = L, R} \sum_{n = 1}^\infty
        \lambda_{\alpha, n} \hbar \omega_n (a_n + a_n^\dagger) n_\alpha,
\end{eqnarray}
where $a_n$ ($a_n^\dagger$) is the annihilation (creation)
operator for the phonon with wavenumber
$q_n = n \pi / \ell$ $(n = 1, 2, 3, \ldots)$.
The phonon energy is given by
\begin{eqnarray}
    \hbar \omega_n = \hbar v q_n = n \frac{\pi \hbar v}{\ell},
\end{eqnarray}
using sound velocity $v$, for small $n$'s.
The dimensionless coupling constants are
\begin{eqnarray}
    \lambda_{\alpha, n} \simeq \frac{3}{\sqrt{n (\ell_\perp / {\rm nm})}}
        \cos (q_n x_\alpha),
\end{eqnarray}
with $(\ell_\perp / {\rm nm})$ being the circumference of
the CNT in units of nanometer
    \cite{Mariani2009}.
When $x_L$ and $x_R$ are symmetric with respect to $x = \ell/2$,
$\lambda_{L, 1} = - \lambda_{R, 1}$.
Disregarding the higher modes of $n \geq 2$,
we obtain the effective Hamiltonian in equation (\ref{eq:H_eff})
with $a_A = a_1$, $\hbar \omega_{\rm ph} = \hbar \omega_1$,
$\lambda_A = \lambda_{L, 1}$, and $\lambda_S = 0$.

\section{Analytic expression for $A$-phonon distribution at current peaks
    \label{app:A-phonon}}

In this appendix,
we derive analytical expressions for the current $I$,
number of phonons $\langle N_A \rangle$, and
autocorrelation function of phonons $g^{(2)}_A (0)$
in equations (\ref{eq:I_A})--(\ref{eq:g2_A}) when
the level spacing $\Delta$ is tuned to the current subpeaks in
figure \ref{fig:Delta}; $\Delta = \Delta_{\nu}
(\simeq \nu \hbar \omega_{\rm ph})$.
We assume that $\lambda_S = 0$ and consider the large bias-voltage
limit. The energy eigenstates are given by the zero-electron states
$
    |0, n \rangle_{eA} = |0 \rangle_e \otimes |n \rangle_A,
$
bonding and anti-bonding states between the polarons
$
    |\pm, n \rangle_{eA} = \frac{1}{\sqrt{2}} \left( |L, n \rangle_{eA}
        \pm |R, n + \nu \rangle_{eA} \right),
$
and polarons localized in dot $R$,
$
    |R, n \rangle_{eA}
$
$
    (n = 0, 1, 2, \ldots, \nu - 1),
$
in a good approximation for $V_{\rm C} \ll \hbar \omega_{\rm ph}$,
as mentioned in section \ref{sec:A-phonon}.
$|L, n \rangle_{eA}$ and $|R, n \rangle_{eA}$ are given in
equation (\ref{eq:polaronA}). The density matrix is given by
\begin{eqnarray*}
\fl \rho_{eA} = \sum_{n = 0}^\infty P_{0, n} |0, n \rangle_{eA~eA} \langle 0, n|
        + \sum_{\sigma = \pm} \sum_{n = 0}^\infty
            P_{\sigma, n} |\sigma, n \rangle_{eA~eA} \langle \sigma, n|
        + \sum_{n = 0}^{\nu - 1} P_{R, n} |R, n \rangle_{eA~eA} \langle R, n|
\end{eqnarray*}
in the BMS approximation.
The occupation numbers for zero-electron states, $n_0$,
bonding or anti-bonding states between the polarons, $n_{\rm mol}$,
and polarons localized in dot $R$, $n_R$, are given by
\begin{eqnarray*}
    n_0 &= \sum_{n = 0}^\infty |0, n \rangle_{eA~eA} \langle 0, n|
        = |0 \rangle_{e~e} \langle 0|,
        \\
    n_{\rm mol} &= \sum_{\sigma = \pm} \sum_{n = 0}^\infty
        |\sigma, n \rangle_{eA~eA} \langle \sigma, n|,
        \\
    \tilde n_R &= \sum_{n = 0}^{\nu - 1} |R, n \rangle_{eA~eA} \langle R, n|,
\end{eqnarray*}
respectively. They satisfy the relation of $n_0 + n_{\rm mol} + \tilde n_R=1$.
The electron number in the DQD is given by
$n_e = n_{\rm mol} + \tilde n_R = 1 - n_0$.
The expectation values of these occupation numbers are expressed as
\begin{eqnarray*}
    \langle n_0 \rangle = \sum_{n = 0}^\infty P_{0, n},
        \qquad
    \langle n_{\rm mol} \rangle = \sum_{n = 0}^\infty P_{{\rm mol}, n},
        \qquad
    \langle \tilde n_R \rangle = \sum_{n = 0}^{\nu - 1} P_{R, n}.
\end{eqnarray*}

In the stationary state,
the equations (\ref{eq:rate_0_A})--(\ref{eq:rate_R_A}) yield
\begin{eqnarray}
\fl 0 = -\Gamma_L P_{0, n}
        + \sum_{m = 0}^{\infty} \frac{\Gamma_R}{2}
        |_A \langle n | \mathcal{T}_A^\dagger | m + \nu \rangle_A|^2
        P_{{\rm mol}, m} + \sum_{m = 0}^{\nu - 1} \Gamma_R
        |_A \langle n | \mathcal{T}_A^\dagger | m \rangle_A|^2 P_{R, m}
        \nonumber \\
    + \Gamma_{\rm ph} \bigl[ (n + 1) P_{0, n + 1} - n P_{0, n} \bigr],
        \label{eq:app_rate_0} \\
\fl 0 = - \frac{\Gamma_R}{2} P_{{\rm mol}, n}
        + \sum_{m = 0}^{\infty} \Gamma_L
        |_A \langle n | \mathcal{T}_A | m \rangle_A|^2 P_{0, m}
        \nonumber \\
    + \Gamma_{\rm ph} \left [
        \left( n + 1 + \frac{\nu}{2} \right) P_{{\rm mol}, n + 1}
        - \left( n + \frac{\nu}{2} \right)
        P_{{\rm mol}, n} \right]
        \label{eq:app_rate_mol}
\end{eqnarray}
with $P_{{\rm mol}, n} = P_{+, n} + P_{-, n}$ $(n = 0, 1, 2, \ldots)$, and
\begin{eqnarray}
    0 &= - \Gamma_R P_{R, n}
        + \Gamma_{\rm ph} \bigl[ (n + 1) P_{R, n + 1} - n P_{R, n} \bigr],
        \label{eq:app_rate_R}
\end{eqnarray}
with $P_{R, \nu} = P_{{\rm mol}, 0} / 2$
$(n = 0, 1, 2, \ldots, \nu - 1)$.

\subsection{Current and electron number}

First, we calculate the current
$I = e \Gamma_L \langle n_0 \rangle$. For the purpose,
we sum up both sides of equation (\ref{eq:app_rate_0}) over $n$.
Using
\begin{eqnarray*}
    \sum_{n = 0}^\infty |_A
        \langle n | \mathcal{T}_A^\dagger
        | m \rangle_A |^2
    &= \left._A \langle m | \mathcal{T}_A
        \left( \sum_n |n \rangle_{AA} \langle n| \right)
        \mathcal{T}_A^\dagger | m \rangle_A \right.
    = 1,
\end{eqnarray*}
we obtain
\begin{eqnarray*}
    0 = -\Gamma_L \langle n_0 \rangle
        + \frac{\Gamma_R}{2} \langle n_{\rm mol} \rangle
        + \Gamma_R \langle \tilde n_R \rangle.
\end{eqnarray*}
Since
$\langle n_0 \rangle + \langle n_{\rm mol} \rangle
+ \langle \tilde n_R \rangle=1$,
we obtain
\begin{eqnarray*}
    \langle n_0 \rangle = \frac{\gamma}{2 + \gamma}
        (1 + \langle \tilde n_R \rangle),
        \qquad
    \langle n_{\rm mol} \rangle = \frac{2}{2 + \gamma}
        \left[ 1 - (1 + \gamma) \langle \tilde n_R \rangle \right],
\end{eqnarray*}
where $\gamma = \Gamma_R / \Gamma_L$.
These equations result in equation (\ref{eq:I_A}), i.e.,
\begin{eqnarray*}
    I = e \Gamma_R \frac{1 + \langle \tilde n_R \rangle}{2 + \gamma}, \qquad
    \langle n_e \rangle = \frac{2 - \gamma \langle \tilde n_R \rangle}
        {2 + \gamma}.
\end{eqnarray*}
The summation of equation (\ref{eq:app_rate_R}) over $n$ yields
\begin{eqnarray}
    \langle \tilde n_R \rangle = \frac{\nu \Gamma_{\rm ph}}{2 \Gamma_R}
        P_{{\rm mol}, 0}.
        \label{eq:app_n_R}
\end{eqnarray}

\subsection{Phonon number}

Next, we derive the phonon number
\begin{eqnarray*}
\langle N_A \rangle
    &= \sum_{n = 0}^\infty P_{0, n}
        ~_{eA} \langle 0, n| N_A |0, n \rangle_{eA}
        + \sum_{\sigma = \pm} \sum_{n = 0}^\infty P_{\sigma, n}
        ~_{eA} \langle \sigma, n | N_A |\sigma, n \rangle_{eA} \\
& \qquad + \sum_{n = 0}^{\nu - 1} P_{R, n}
        ~_{eA} \langle R, n| N_A |R, n \rangle_{eA} \\
&= \sum_{n = 0}^\infty n P_{0, n} + \sum_{n = 0}^\infty
        \left( n + \frac{\nu}{2} + \lambda_A^2 \right) P_{{\rm mol}, n}
        + \sum_{n = 0}^{\nu - 1} (n + \lambda_A^2) P_{R, n} \\
&\equiv \langle N_A n_0 \rangle
        + \langle N_A n_{\rm mol} \rangle + \langle N_A \tilde n_R \rangle.
\end{eqnarray*}
We have used the relation,
$
    \mathcal{T}_A^\dagger N_A \mathcal{T}_A
        = ( \mathcal{T}_A^\dagger a_A^\dagger \mathcal{T}_A )
            ( \mathcal{T}_A^\dagger a_A \mathcal{T}_A )
        = (a_A^\dagger - \lambda_A) (a_A - \lambda_A)
$.
We multiply both sides of equations (\ref{eq:app_rate_0})--(\ref{eq:app_rate_R})
by $n$ and sum up over $n$. Then we find
\begin{eqnarray}
    0 &= - (\Gamma_L + \Gamma_{\rm ph}) \langle N_A n_0 \rangle
        + \frac{\Gamma_R}{2} \langle N_A n_{\rm mol} \rangle
        + \Gamma_R \langle N_A \tilde n_R \rangle
        + \frac{\nu \Gamma_R}{4} \langle n_{\rm mol} \rangle,
        \label{eq:app_N_A_0} \\
    0 &= \Gamma_L \langle N_A n_0 \rangle
        - \left( \frac{\Gamma_R}{2} +\Gamma_{\rm ph} \right)
            \langle N_A n_{\rm mol} \rangle
        + \lambda_A^2 \Gamma_L \langle n_0 \rangle
        \nonumber \\
    &\qquad + \left( \frac{\nu + 2 \lambda_A^2}{4} \Gamma_R
            + \lambda_A^2 \Gamma_{\rm ph} \right) \langle n_{\rm mol} \rangle
        + \Gamma_R \langle \tilde n_R \rangle,
        \label{eq:app_N_A_mol} \\
    0 &= - (\Gamma_R + \Gamma_{\rm ph}) \langle N_A \tilde n_R \rangle
    + \left[ (\nu - 1 + \lambda_A^2) \Gamma_R
        + \lambda_A^2 \Gamma_{\rm ph} \right] \langle \tilde n_R \rangle.
        \label{eq:app_N_A_R}
\end{eqnarray}
Here, we have used
\begin{eqnarray}
    \sum_{n = 0}^\infty n |_A
        \langle n | \mathcal{T}_A^\dagger
        | m \rangle_A |^2
    &= \left._A \langle m | \mathcal{T}_A
        N_A \left( \sum_n |n \rangle_{AA} \langle n| \right)
        \mathcal{T}_A^\dagger | m \rangle_A \right.
        \nonumber \\
    &= \sum_{m} \left._A \langle m | \mathcal{T}_A N_A \mathcal{T}_A^\dagger
    | m \rangle_A \right..
    \label{eq:app_tech}
\end{eqnarray}
From equations (\ref{eq:app_N_A_0})--(\ref{eq:app_N_A_R}), we obtain
equation (\ref{eq:N_A}), i.e.,
\begin{eqnarray*}
    \langle N_A \rangle = (\nu + 2 \lambda_A^2) \frac{I}{e \Gamma_{\rm ph}}
        + \lambda_A^2 \langle n_e \rangle.
\end{eqnarray*}

\subsection{Phonon autocorrelation function}

Finally, we derive the phonon autocorrelation function at $\tau=0$,
\begin{eqnarray*}
    g^{(2)}_A (0) = \frac{\langle :N_A^2: \rangle}{\langle N_A \rangle^2}
        = \frac{\langle N_A^2 \rangle - \langle N_A \rangle}
            {\langle N_A \rangle^2},
\end{eqnarray*}
where
\begin{eqnarray*}
\langle N_A^2 \rangle
    &= \sum_{n = 0}^\infty P_{0, n}
        ~_{eA} \langle 0, n| N_A^2 |0, n \rangle_{eA}
        + \sum_{\sigma = \pm} \sum_{n = 0}^\infty P_{\sigma, n}
        ~_{eA} \langle \sigma, n | N_A^2 |\sigma, n \rangle_{eA} \\
& \qquad + \sum_{n = 0}^{\nu - 1} P_{R, n}
        ~_{eA} \langle R, n| N_A^2 |R, n \rangle_{eA} \\
&= \sum_{n = 0}^\infty n^2 P_{0, n} + \sum_{n = 0}^\infty
        \left[ n^2 + \lambda_A^2 (4n + 1 + \lambda_A^2)
        + \nu \left( n + \frac{\nu}{2} + 2 \lambda_A^2 \right) \right]
        P_{{\rm mol}, n}
        \nonumber \\
& \qquad + \sum_{n = 0}^{\nu - 1}
        \left[ n^2 + \lambda_A^2 (4 n + 1 + \lambda_A^2) \right]
        P_{R, n} \\
&\equiv \langle N_A^2 n_0 \rangle
        + \langle N_A^2 n_{\rm mol} \rangle + \langle N_A^2 \tilde n_R \rangle.
\end{eqnarray*}
We multiply both sides of equations (\ref{eq:app_rate_0})--(\ref{eq:app_rate_R})
by $n^2$ and sum up over $n$.
A similar technique to equation (\ref{eq:app_tech}) leads to
\begin{eqnarray}
\fl 0 = -(\Gamma_L + 2 \Gamma_{\rm ph}) \langle N_A^2 n_0 \rangle
        + \frac{\Gamma_R}{2} \langle N_A^2 n_{\rm mol} \rangle
        + \Gamma_R \langle N_A^2 \tilde n_R \rangle
        + \Gamma_{\rm ph} \langle N_A n_0 \rangle
        \nonumber \\
    + \frac{\nu \Gamma_R}{2} \langle N_A n_{\rm mol} \rangle
        + \frac{\nu \lambda_A^2 \Gamma_R}{2} \langle n_{\rm mol} \rangle,
        \label{eq:app_N_A_2_0} \\
\fl 0 = \Gamma_L \langle N_A^2 n_0 \rangle
        - \left( \frac{\Gamma_R}{2} + 2 \Gamma_{\rm ph} \right)
        \langle N_A^2 n_{\rm mol} \rangle
        + 4 \lambda_A^2 \Gamma_L \langle N_A n_0 \rangle
        \nonumber \\
    + \left[ \frac{\nu + 4 \lambda_A^2}{2} \Gamma_R
            + ( \nu + 1 + 8 \lambda_A^2) \Gamma_{\rm ph} \right]
            \langle N_A n_{\rm mol} \rangle
            \lambda_A^2 (1 + \lambda_A^2) \Gamma_L \langle n_0 \rangle
        \nonumber \\
    + \lambda_A^2 \left[ \frac{1 - \nu - 3 \lambda_A^2}{2} \Gamma_R
            + (1 - \nu - 6 \lambda_A^2) \Gamma_{\rm ph} \right]
            \langle n_{\rm mol} \rangle
        - \Gamma_R \langle \tilde n_R \rangle,
        \label{eq:app_N_A_2_mol} \\
\fl 0 = -(\Gamma_R + 2 \Gamma_{\rm ph}) \langle N_A^2 \tilde n_R \rangle
        + \left[ 4 \lambda_A^2 \Gamma_R
            + (1 + 8 \lambda_A^2) \Gamma_{\rm ph} \right]
            \langle N_A \tilde n_R \rangle
        \nonumber \\
    + \left\{ \left[ (\nu - 1)^2 + \lambda_A^2 (1 - 3\lambda_A^2) \right]
                \Gamma_R
            + \lambda_A^2 ( 1 - 6 \lambda_A^2) \Gamma_{\rm ph} \right\}
            \langle \tilde n_R \rangle
        \label{eq:app_N_A_2_R}.
\end{eqnarray}
From equations (\ref{eq:app_N_A_2_0})--(\ref{eq:app_N_A_2_R}), we find
\begin{eqnarray*}
\fl \langle N_A^2 \rangle - \langle N_A \rangle
    = 2 \lambda_A^2 \frac{\Gamma_L}{\Gamma_{\rm ph}} \langle N_A n_0 \rangle
        + \left( \frac{\nu + 2 \lambda_A^2}{2} \frac{\Gamma_R}{\Gamma_{\rm ph}}
            + \frac{\nu + 4 \lambda_A^2}{2} \right)
            \langle N_A n_{\rm mol} \rangle
        \nonumber \\
    + 2 \lambda_A^2 \left( \frac{\Gamma_R}{\Gamma_{\rm ph}} + 2 \right)
        \langle N_A \tilde n_R \rangle
        - \frac{\nu + 2 \lambda_A^4}{2} \frac{\Gamma_L}{\Gamma_{\rm ph}}
            \langle n_0 \rangle
        \nonumber \\
    - \frac{\lambda_A^2 (\nu + 6 \lambda_A^2)}{2}
            \langle n_{\rm mol} \rangle
        - \left[ \frac{\nu (2 - \nu)}{2} \frac{\Gamma_R}{\Gamma_{\rm ph}}
            + 3 \lambda_A^2 \right] \langle \tilde n_R \rangle.
\end{eqnarray*}

Now we evaluate $g^{(2)}_A (0)$ when $\Gamma_{L, R} \gg \Gamma_{\rm ph}$.
In this case,
$
    \langle \tilde n_R \rangle = \mathcal{O} (\Gamma_{\rm ph} / \Gamma_{L, R})
$
from equation (\ref{eq:app_n_R}). Then
\begin{eqnarray*}
    I = \frac{e \Gamma_R}{2 + \gamma}
        + \mathcal{O}(\Gamma_{\rm ph} / \Gamma_{L, R}),
        \qquad
    \langle N_A \rangle
    = \frac{\nu + 2 \lambda_A^2}{2 + \gamma}
                \frac{\Gamma_R}{\Gamma_{\rm ph}}
        + \mathcal{O}(1).
\end{eqnarray*}
Equations (\ref{eq:app_N_A_0}) and (\ref{eq:app_N_A_R}) yield
\begin{eqnarray*}
    2 \langle N_A n_0 \rangle
        = \gamma \langle N_A n_{\rm mol} \rangle
            + \mathcal{O} (1), \qquad
    \langle N_A \tilde n_R \rangle
        = \mathcal{O}(\Gamma_{\rm ph} / \Gamma_{L, R}).
\end{eqnarray*}
Using
$
    \langle N_A \rangle
    = \langle N_A n_0 \rangle
        + \langle N_A n_{\rm mol} \rangle
        + \langle N_A \tilde n_R \rangle
$, we have
\begin{eqnarray*}
    \langle N_A n_0 \rangle
        = (\nu + 2 \lambda_A^2) \frac{\gamma}{(2 + \gamma)^2}
            \frac{\Gamma_R}{\Gamma_{\rm ph}} + \mathcal{O} (1),
        \\
    \langle N_A n_{\rm mol} \rangle
        = (\nu + 2 \lambda_A^2) \frac{2}{(2 + \gamma)^2}
                    \frac{\Gamma_R}{\Gamma_{\rm ph}} + \mathcal{O}(1).
\end{eqnarray*}
Using these relations, we obtain equation (\ref{eq:g2_A}), i.e.,
\begin{eqnarray*}
    g^{(2)}_A (0) = \frac{\nu + 4 \lambda_A^2}{\nu + 2 \lambda_A}
        + \mathcal{O} (\Gamma_{\rm ph} / \Gamma_{L, R}).
\end{eqnarray*}

\subsection{Case of large decay rate of phonon
    \label{app:large-decay-phonon}}

Here, we comment on the opposite limit of
$\Gamma_{L, R} \ll \Gamma_{\rm ph}$.
The analytical expressions for the current, $g^{(2)}_A(0)$, etc.,
can be obtained in a similar way to the case of
$\Gamma_{L, R} \gg \Gamma_{\rm ph}$.
At the main peak of the current ($\Delta = 0$), the electron
number and current are written as
\begin{eqnarray*}
    \langle n_e \rangle
    = \langle n_{\rm mol} \rangle \simeq \frac{1}{2 + \gamma},
    \qquad
    I \simeq \frac{e \Gamma_R}{2 + \gamma},
\end{eqnarray*}
whereas the phonon number and its autocorrelation function are
\begin{eqnarray*}
    \langle N_A \rangle \simeq \frac{\lambda_A^2}{2 + \gamma},
    \qquad g^{(2)}_A(0) \simeq 2 + \gamma.
\end{eqnarray*}
At the $\nu$th subpeak of the current
($\Delta = \Delta_{\nu} \simeq \nu \hbar \omega_{\rm ph}$), we obtain
\begin{eqnarray*}
    \langle n_e \rangle \simeq \langle \tilde n_R \rangle
    \simeq \frac{1}{1 + \gamma},
    \qquad \langle n_{\rm mol} \rangle \simeq 0,
    \qquad I \simeq \frac{e \Gamma_R}{1 + \gamma},
\end{eqnarray*}
and
\begin{eqnarray}
    \langle N_A \rangle \simeq \frac{\lambda_A^2}{1 + \gamma},
    \qquad g^{(2)}_A(0) \simeq 1 + \gamma.
    \label{eq:app_g2_slow_tunneling}
\end{eqnarray}
As discussed in section \ref{subsec:antibunching},
the phonon bunching is observed even for
small $\lambda_A$ in the case of $\Gamma_{L, R} \ll \Gamma_{\rm ph}$.
In this case, an electron is localized in dot $R$ for a long time,
forming a polaron $|R,0 \rangle_{eA}$, after a phonon is immediately decayed.
Thus the number of phonons created by the interdot tunneling is much
smaller than that accompanied by the polaron staying in dot $R$.
This situation results in the bunched phonons.

\section{Rate equation for $S$-phonon
    \label{app:S-phonon}}

In this appendix, we derive the rate equation in the presence of
$S$-phonons only ($\lambda_A = 0$).
We can analytically diagonalize Hamiltonian $H$ for any $\Delta$
in this case.
The eigenstates of $H$ are given by the zero-electron states,
$|0, n \rangle_{eS} = |0 \rangle_e \otimes |n \rangle_S$,
and bonding and anti-bonding states between the polarons,
\begin{eqnarray}
    |+, n \rangle_{eS} &= \cos \frac{\theta}{2} |L, n \rangle_{eS}
         + \sin \frac{\theta}{2} |R, n \rangle_{eS},
    \label{eq:polaronSplus} \\
    |-, n \rangle_{eS} &= - \sin \frac{\theta}{2} |L, n \rangle_{eS}
         + \cos \frac{\theta}{2} |R, n \rangle_{eS},
    \label{eq:polaronSminus}
\end{eqnarray}
with $\tan \theta = 2 V_{\rm C} / \Delta$.
$|L, n \rangle_{eS}$ and $|R, n \rangle_{eS}$ are given in
equation (\ref{eq:polaronS}).
The corresponding energy eigenvalues are
$\epsilon_{0, n} = n \hbar \omega_{\rm ph}$
and
\begin{eqnarray*}
    \epsilon_{\pm, n} &= \pm \left[ (\Delta/2)^2 + V_{\rm C}^2 \right]^{1/2}
        + (n - \lambda_S^2) \hbar \omega_{\rm ph}
\end{eqnarray*}
$(n = 0, 1, 2, \ldots)$, respectively.
Using the dissipator $\mathcal{L}_e$ in equation (\ref{eq:L_e}),
we obtain the rate equations under a finite bias voltage as
\begin{eqnarray}
\fl \dot{P}_{0, n} = & - \left[ \sum_{\alpha = L, R} \sum_{\sigma = \pm}
        \sum_{m = 0}^\infty
        f_\alpha(\epsilon_{\sigma, m} - \epsilon_{0, n})
        \Gamma_{\alpha, \sigma}
        |_S \langle m | \mathcal{T}_S^\dagger | n \rangle_S|^2
        \right] P_{0, n} \nonumber \\
\fl & + \sum_{\alpha, \sigma, m}
        \bar{f}_\alpha(\epsilon_{\sigma, m} - \epsilon_{0, n})
        \Gamma_{\alpha, \sigma}
        |_S \langle n | \mathcal{T}_S | m \rangle_S|^2
        P_{\sigma, m}
        + \Gamma_{\rm ph} \bigl[ (n + 1) P_{0, n + 1}
        - n P_{0, n} \bigr],
        \label{eq:app_rate_S_0} \\
\fl \dot{P}_{\pm, n} = & - \left[ \sum_{\alpha, m}
        \bar{f}_\alpha (\epsilon_{\pm, n} - \epsilon_{0, m})
        \Gamma_{\alpha, \pm} |_S \langle m | \mathcal{T}_S | n \rangle_S|^2
        \right] P_{\pm, n} \nonumber \\
\fl & + \sum_{\alpha, m} f_\alpha (\epsilon_{\pm, n} - \epsilon_{0, m})
        \Gamma_{\alpha, \pm}
        |_S \langle n | \mathcal{T}_S^\dagger | m \rangle_S|^2
        P_{0, m}
        + \Gamma_{\rm ph} \bigl[ (n + 1) P_{\pm, n + 1}
        - n P_{\pm, n} \bigr].
        \label{eq:app_rate_S_pm}
\end{eqnarray}
Here, we have introduced tunnel coupling strength
$\Gamma_{\alpha, \sigma}$ between lead $\alpha$ and molecule orbital
$\sigma$ in equations (\ref{eq:polaronSplus}) and
(\ref{eq:polaronSminus}):
$\Gamma_{L, +} = \Gamma_L \cos^2 (\theta/2)$,
$\Gamma_{L, -} = \Gamma_L \sin^2 (\theta/2)$,
$\Gamma_{R, +} = \Gamma_R \sin^2 (\theta/2)$, and
$\Gamma_{R, -} = \Gamma_R \cos^2 (\theta/2)$.
In the limit of large bias, these equations yield
\begin{eqnarray}
\fl \dot{P}_{0, n} &= - \Gamma_L P_{0, n}
        + \sum_{\sigma = \pm} \sum_{m = 0}^\infty \Gamma_{R, \sigma}
            |_S \langle n | \mathcal{T}_S | m \rangle_S|^2 P_{\sigma, m}
        + \Gamma_{\rm ph} \bigl[ (n + 1) P_{0, n + 1}
            - n P_{0, n} \bigr],
            \label{eq:rate_0_S_large_V} \\
\fl \dot{P}_{\pm, n} &= - \Gamma_{R, \pm} P_{\pm, n}
        + \sum_{m = 0}^\infty \Gamma_{L, \pm}
            |_S \langle n | \mathcal{T}_S^\dagger | m \rangle_S |^2
            P_{0, m}
        + \Gamma_{\rm ph} \bigl[ (n + 1) P_{\pm, n + 1}
            - n P_{\pm, n} \bigr].
            \label{eq:rate_pm_S_large_V}
\end{eqnarray}
We sum up both sides of equations (\ref{eq:rate_0_S_large_V})
and (\ref{eq:rate_pm_S_large_V}) over $n$ to obtain
the reduced rate equations for electrons,
\begin{eqnarray}
    \dot{P}_0 &= - \Gamma_L P_0
        + \sum_{\sigma = \pm} \Gamma_{R, \sigma} P_{\sigma},
        \label{eq:rate_0_e} \\
    \dot{P}_{\pm} &= - \Gamma_{R,\pm} P_{\pm} + \Gamma_{L,\pm} P_0,
        \label{eq:rate_pm_e}
\end{eqnarray}
with $P_0 = \sum_{n = 0}^\infty P_{0, n}$ and
$P_\pm = \sum_n P_{\pm, n}$.
The equations (\ref{eq:rate_0_e}) and (\ref{eq:rate_pm_e})
are the same as those in the absence of electron-phonon coupling.
This indicates that $S$-phonons do not affect the electron transfer
in the large bias-voltage limit, whereas they are created by
the current through the Franck-Condon effect.
The further calculation yields
the current, electron and $S$-phonon numbers, and
phonon autocorrelation function in equations
(\ref{eq:I_S}), (\ref{eq:N_S}), and (\ref{eq:g2_S}),
in a similar way to \ref{app:A-phonon}.

\section*{References}


\providecommand{\newblock}{}

\end{document}